\documentclass[sigconf]{acmart}
\usepackage{xurl}
\usepackage{multirow}

\AtBeginDocument{%
  \urlstyle{tt}%
}

\makeatletter
\def\@acmConferenceHeader{}
\makeatother

\setcopyright{none}
\renewcommand\footnotetextcopyrightpermission[1]{}
\AtBeginDocument{%
  \pagestyle{empty}%
  \thispagestyle{empty}%
}

\AtBeginDocument{%
  }

\copyrightyear{2026}
\acmYear{2026}
\acmDOI{XXXXXXX.XXXXXXX}
\acmConference[]{}{}{}

\begin{document}

\title{SF-LIFE: A Large-Scale Simulated Movement Dataset for the San Francisco Bay Area}

\author{Chanuka Algama}
\affiliation{%
  \institution{Tulane University}
  \country{USA}
}

\author{Taylor Anderson}
\affiliation{%
\institution{George Mason University}
\country{USA}
}

\author{Henrique Ferraz de Arruda}
\affiliation{%
  \institution{BIFI University of Zaragoza,\\ARAID Foundation}
  \country{Spain}
}

\author{Andrew Crooks}
\affiliation{%
\institution{University at Buffalo}
\country{USA}
}

\author{Nathan Holt}
\affiliation{%
  \institution{L3Harris Technologies}
  \country{USA}
}

\author{Erfan Hosseini Sereshgi}
\affiliation{%
  \institution{Tulane University}
  \country{USA}
}

\author{John Hunter}
\affiliation{%
  \institution{L3Harris Technologies}
  \country{USA}
}

\author{Hamdi Kavak}
\affiliation{%
\institution{George Mason University}
\country{USA}
}

\author{Lance Kennedy}
\affiliation{%
  \institution{Emory University}
  \country{USA}
}

\author{Yueyang Liu}
\affiliation{%
  \institution{Emory University}
  \country{USA}
}

\author{Dieter Pfoser}
\affiliation{%
  \institution{George Mason University}
  \country{USA}
}

\author{Sandro Martinelli Reia}
\affiliation{%
  \institution{George Mason University}
  \country{USA}
}

\author{Doug Taylor}
\affiliation{%
  \institution{L3Harris Technologies}
  \country{USA}
}

\author{Mauryan Uppalapati}
\affiliation{%
  \institution{Tulane University}
  \country{USA}
}

\author{Boyu Wang}
\affiliation{%
  \institution{University at Buffalo}
  \country{USA}
}

\author{Carola Wenk}
\affiliation{%
  \institution{Tulane University}
  \country{USA}
}

\author{Andreas Z\"ufle}
\affiliation{%
  \institution{Emory University}
  \country{USA}
}

\renewcommand{\shortauthors}{Your Name}

\begin{abstract}
  We introduce SF-LIFE, a large-scale simulated movement dataset designed to accelerate research in transportation, mobility, and machine learning. The dataset contains 3,024,000,000,000 location records capturing complete, noise-free, multi-modality trajectories of $500,000$ simulated agents observed at a 1Hz frequency navigating the San Francisco Bay Area network over a 70-day period. The data captures (1) needs-driven daily agendas of individual agents generated by an agent-based simulation of human patterns of life and (2) detailed kinematic trajectories moving agents across the OpenStreetMap representation of San Francisco using data from 40+ transit agencies across 9 counties.
  SF-LIFE provides unprecedented scale and detail as trajectories are based on real transit infrastructure using San Francisco General
Transit Feed Specification (GTFS) data, having agent movements across multiple modalities, including bus, rail, bike, automobile, and walking. For this high-fidelity simulated representation of San Francisco, we provide (1) the full trajectory data annotated with transportation mode labels, (2) reduced-size versions of the trajectory data with reduced temporal frequency, (3) agent activity information describing the causal activity why an agent visits a place, (4) agent demographic data, and (5) the underlying OSM road network and building data.

As the first dataset of its scale and level of detail, SF-LIFE overcomes the privacy, noise, and completeness limitations inherent in real-world tracking data, providing a robust and ethically sourced resource for research in transit optimization, human mobility analysis, and urban computing.
\end{abstract}


\maketitle

\section{Introduction}
Urban mobility research and transportation planning increasingly rely on large-scale movement datasets to understand human behavior patterns, optimize transit systems, and develop intelligent transportation solutions \cite{urban_mobility_analysis, trajectory_mining}. 
However, real-world movement data is often noisy, incomplete, and subject to privacy constraints, making it difficult to design robust algorithms and models \cite{privacy_mobility}. To address these challenges, we present SF-LIFE, a comprehensive simulated movement dataset for the San Francisco Bay Area that provides clean, complete trajectory data for $500,000$ agents over a 70-day period.

The SF-LIFE dataset represents a significant advancement in spatial data analysis \cite{spatial_data_analysis}, offering unprecedented scale and complexity for transportation research. With 3 trillion location trajectory records capture the location of 500,000 agents at a 1Hz frequency (one location per second) over a period of 70 Days. The data captures (1) realistic daily agendas generated by a agent-based simulation of human patterns of life and (2) detailed kinematic trajectories moving agents across the OpenStreetMap representation of San Francisco using data from 40+ transit agencies across 9 counties. This provides a unique combination of geographic breadth, temporal depth, and data quality that is currently unavailable in existing movement datasets \cite{movement_patterns, urban_computing, large_scale_mobility}. The integration of GTFS (General Transit Feed Specification) compliant transit infrastructure data with detailed agent trajectory information creates a comprehensive foundation for spatial analytics, machine learning applications, and urban computing research.

This dataset addresses critical gaps in spatial data analysis by providing: 
\begin{enumerate}
    \item complete trajectory coverage without GPS signal loss or device failures,
    \item realistic multi-modal transportation patterns across a complex urban network,
    \item privacy-preserving simulated data that maintains statistical validity, and
    \item standardized data formats that enable reproducible research.
\end{enumerate}

The scale and complexity of SF-LIFE make it particularly valuable for developing and benchmarking spatial analysis algorithms, transportation optimization models, and machine learning approaches for urban mobility.

\section*{Dataset Availability}
The SF-LIFE dataset is publicly available under the ODC-By license at \url{https://huggingface.co/datasets/sf-life/sf-life}.

\section{Related Work}

Movement datasets have become increasingly important for transportation research and urban analytics \cite{trajectory_mining, urban_computing}. Previous work includes GPS tracking studies \cite{movement_patterns}, mobile phone data analysis, GTFS feeds \cite{gtfs_spec}, and automated passenger counting systems. 
A related line of work has focused on agent-based and activity-based models for generating synthetic human mobility and patterns-of-life data. One approach introduced \textit{Urban Life}, a model of people and places designed to simulate urban activity patterns through interactions between individuals and the built environment~\cite{zufle2023urban}. Subsequent work proposed a patterns-of-life human mobility simulation framework that generates individual-level mobility traces from behavioral routines and activity constraints~\cite{amiri2024patterns}. More recently, HD-GEN was presented as a software system for the generation of human mobility data based on patterns of life, with a particular emphasis on the necessity of producing synthetic mobility data on a large scale~\cite{amiri2025hd}.
However, existing datasets suffer from significant limitations that hinder spatial data analysis research.

\subsection{Limitations of Existing Datasets}

Current movement datasets face several critical challenges: (1) \textbf{Scale limitations} - most datasets cover fewer than $100,000$ individuals over limited time periods, (2) \textbf{Geographic constraints} - coverage is often limited to single cities or regions, (3) \textbf{Data quality issues} - GPS signal loss, device failures, and user opt-outs create incomplete trajectories, (4) \textbf{Privacy concerns} - real-world tracking data raises ethical and legal issues \cite{privacy_mobility}, and (5) \textbf{Modal limitations} - most datasets focus on single transportation modes rather than multi-modal networks.

\subsection{Advantages of SF-LIFE}

SF-LIFE addresses these limitations through its unprecedented scale and complexity. The dataset's 500,000 agents represent the largest simulated population in transportation research, providing statistical power for spatial analysis that is unavailable in existing datasets \cite{agent_based_simulation}. The integration of 40+ transit agencies across 9 counties creates a realistic multi-modal transportation network that mirrors the complexity of real urban systems \cite{transit_optimization}.
By integrating high-density agent populations with detailed demographic profiles and real-world building geometries, this dataset offers a comprehensive roadmap for urban mobility research. Its curated trajectory data eliminates the noise and sparsity typical of raw observations, allowing researchers to refine spatial analysis algorithms with precision. Furthermore, the privacy-preserving simulation framework \cite{privacy_mobility} ensures that realistic movement patterns are maintained for open-source benchmarking without compromising data ethics.

\subsection{Impact on Spatial Data Analysis}

SF-LIFE's scale and complexity make it particularly valuable for spatial data analysis research. The dataset enables: (1) \textbf{Large-scale spatial clustering} - identifying movement patterns across diverse geographic regions, (2) \textbf{Multi-modal network analysis} - studying interactions between different transportation modes, (3) \textbf{Temporal-spatial modeling} - understanding how movement patterns evolve over time and space, and (4) \textbf{Machine learning applications} - training models for trajectory prediction and anomaly detection \cite{machine_learning_mobility}.
The dataset's standardized format and comprehensive documentation facilitate its use as a benchmark for spatial analysis algorithms, enabling fair comparisons between different approaches and promoting reproducible research in the field.

\begin{figure*}[h]
    \centering
    \includegraphics[width=\linewidth]{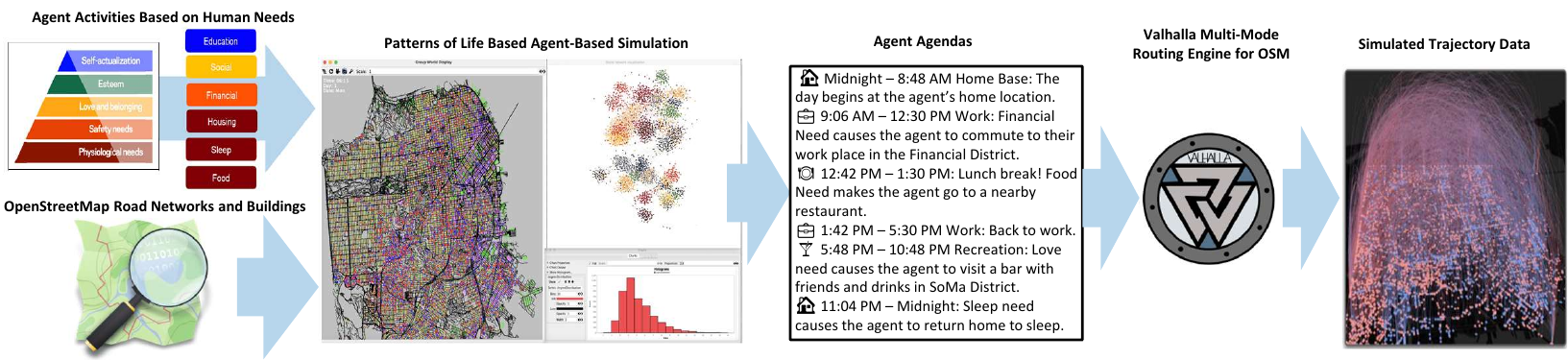}
    \caption{Simulation Architecture: An agent-based city-level simulation uses Maslowian Needs~\cite{maslow1943theory} to create realistic human patterns of life such as going to work and restaurants and meeting friends to satisfy their needs. OpenStreetMap (OSM) data is used to create the simulation environment (buildings, road). The agent-based simulation create daily agendas for each agent. The agendas are fed into the Valhalla multi-mode routing engine to create daily trajectories for agents. }
    \label{fig:architecture}
\end{figure*}

\vspace{-0.1cm}
\section{Simulation Architecture}
This section describes the simulation framework used to generate the data.
The overall architecture of the simulation framework is summarized in Figure~\ref{fig:architecture}. The core of the simulation is an agent-based simulation framework in which each simulated agent models an individual living in the San Francisco Bay Area. The simulation environment using building and road network data from OpenStreetMap (OSM) as described in Section~\ref{subsec:environment}. 
The agents do not correspond to specific real-world individuals; rather, they are synthetically generated with demographic characteristics, attributes, home locations, and workplaces derived from census data, as described in Section~\ref{subsec:initialization}.
Once agents have been initialized and the simulation starts, the behavior of agents is based on human needs which lead to emerging patterns of life as described in Section~\ref{subsec:agent_behavior}. Agents in this simulation maintain dynamically evolving social networks capturing their friends and co-workers which agents need to interact with to satisfy their social needs as described in Section~\ref{subsec:POL}. This agent-based simulation framework yields agent agendas which describe, for each simulated day, what each agent plans to do to satisfy their needs. To turn these agendas into complete trajectories, we use the Valhalla multi-mode routing framework to find shortest paths to route agents between buildings on their agendas as described in Section~\ref{subsec:valhalla}.

\vspace{-0.1cm}
\subsection{Simulation Environment}\label{subsec:environment}
The simulation environment is constructed using OpenStreetMap (OSM) foundation data, which informs the geolocation and functionality of the infrastructure. The initialization process ingests three primary datasets: 
\begin{itemize}
    \item \texttt{\textbf{Buildings}}: Buildings define the physical environment, encompassing all structures within the simulation. Each entry includes geolocation coordinates (of the centroid of the building polygon) and a functional category (residential, workplace, education, religion, restaurant, recreation). 
       \item \texttt{\textbf{Roads}}: The preparation of the road networks and mass transit information (bus and rail) is performed beforehand to ensure the output trajectories conform to local constraints such as road locations, speed limits, and bus departure times. The road network is sourced from OSM and converted into a routable set of tiles via the Mjolnir tool provided by Valhalla (\url{https://valhalla.github.io/valhalla/mjolnir/}). 
    \item \texttt{\textbf{Mass Transit Schedules}}:   Mass transit data is sourced from GTFS files provided by 511 SF Bay and is publicly available (\url{https://511.org/open-data/transit}). The GTFS files are processed to create routable tiles via the same processes as the OSM data, with the additional information about the GTFS provided to allow for the OSM road network to be conflated with the mass transit network.
\end{itemize}

    

\subsection{Synthetic Population and Simulation Initialization}\label{subsec:initialization}
To create a synthetic population we follow the approach presented in~\cite{jiang2024large} to give simulated agents realistic demographics, home locations, and work locations based on U.S. Census data. 
The synthetic population is created using Heuristic Synthesis to align individual agents with 2020 U.S. Census tract-level demographics, grouping them into households based on census structures. It includes sociodemographic attributes (age, gender, employment status), vehicle ownership (car/bike), and specific building IDs associated with the agent's residence and mandatory destinations (workplace or school). Home locations are determined by assigning household IDs to specific residential buildings. Agents are mapped to work locations using the U.S. Census Bureau’s Longitudinal Employer-Household Dynamics (LEHD) Origin-Destination Employment Statistics (LODES) dataset~\cite{abowd2004integrated}. This administrative dataset provides aggregate home-to-work flows between census tracts, which the study used to realistically pair an agent's residential tract to a workplace tract based on historical employment data rather than self-reported travel logs. To initialize the social network between agents, we link agents based on (1) shared household, work, school, and daycare locations, and (2) connecting agents based on a spatial version of the Newman-Watts-Strogatz~\cite{watts1998collective,gallagher2023synthetic} small-world synthetic network generation to generate social network connections. More details on this synthetic population generation can be found in~\cite{jiang2024large}.

\subsection{Agent Behavior: Mandatory and Needs-Driven Activities}\label{subsec:agent_behavior}

For agents to decide what to do and where to travel, the simulation uses a needs-driven behavioral framework in which daily schedules emerge from the interplay between mandatory activities, flexible activities, spatial constraints, opening hours, and social interactions~\cite{reia2026towards}. The distinction between mandatory activities and flexible or needs-driven activities follows transportation research on trip chaining and activity-based travel behavior, where daily mobility is commonly organized around fixed obligations, such as work or school, and more discretionary activities that can be inserted around them~\cite{primerano2008defining}. 
The flexible/needs framework is conceptually based on Maslow's hierarchy of needs~\cite{maslow1943theory} and supported by previous agent-based models that use evolving needs to describe agent behavior and generate patterns of life~\cite{zufle2023urban,amiri2024patterns,amiri2025hd}.

Agents are represented as workers, students, or homemakers. Workers and students have work and school, respectively, as mandatory activities, while homemakers do not have a fixed work or school activity. Flexible activities are driven by time-varying needs. In the model, these needs are:

\begin{itemize}
    \item \textbf{Food need:} represents the agent's need to eat. When this need exceeds its threshold, the agent attempts to schedule a trip to a restaurant, subject to time-budget, travel-time, destination-availability, and opening-hour constraints.

    \item \textbf{Recreation/social need:} represents the agent's need for leisure and social interaction. When this need exceeds its threshold, the agent attempts to schedule a trip to a recreational place. This need is socially modulated: when agents are co-located with friends at recreational places, their social satisfaction increases, existing social ties can be reinforced, and new social connections may be formed.

   \item \textbf{Errand need:} represents a catch-all category for discretionary activities and needs not explicitly captured by the simulation, such as shopping, personal tasks, household-related activities, or other routine non-work and non-school purposes. When this need exceeds its threshold, the agent attempts to schedule a trip to an errand destination.
   
\end{itemize}

Destination choice is constrained by the set of locations available to each agent and follows a rank-based probability mechanism, so that higher-ranked destinations are more likely to be selected. This needs-based decision process, together with mandatory work/school routines, generates daily activity schedules and trip chains. The rates at which flexible needs accumulate are behavioral parameters of the model and can be adjusted to improve agreement with empirical mobility patterns from the United States National Household Travel Survey~\cite{bricka2024summary}.

\subsection{Scalable Patterns of Life Simulation}\label{subsec:POL}

To simulate the movement of agents satisfying their mandatory and flexible activities, we use an agent-based simulation model (ABM) introduced in~\cite{reia2026towards}.
This ABM simulates day-to-day activity-driven mobility at the individual level, generating trip chains as agents move between home, work or school, and flexible destinations such as restaurants, recreational places, and errand locations. The framework is relevant because it produces full-population, behaviorally grounded mobility outputs that reproduce key empirical patterns of life observed in travel data, enabling controlled and scalable analyses of urban mobility and access.

A key feature of the ABM is its transferability across urban contexts. Using complementary mobility metrics, including (i) frequencies of trips to activity destinations, (ii) origin-destination flow patterns, and (iii) the distribution of trips per agent, the standard-parameter configuration reproduces observed ``patterns of life'' in most metropolitan areas with similarity scores typically above $0.80$, without extensive city-specific calibration~\cite{reia2026towards}.

Simulations run in parallel using MPI within Repast4Py~\cite{collier2022distributed}. The shared spatial environment is partitioned across MPI ranks, while a global scheduler enforces synchronized time stepping. Agents follow a daily activity cycle in which mandatory activities depend on agent type, and flexible activities emerge from time-varying needs for food, recreation/social interaction, and errands. Mobility unfolds at a 5-minute resolution as agents transition between travel and dwell states, record visited locations, and may experience social reinforcement when co-located with friends at recreational venues. Social interactions may also reinforce existing ties or create new social connections, which can affect future recreation/social decisions. When agents move across spatial partitions, Repast4Py migrates their state between MPI ranks to maintain seamless execution at scale. This distributed design enables full-population metropolitan simulations, including cases exceeding 20 million agents
such as New York City~\cite{reia2026towards}.

The simulation yields, for each agent, a daily activity schedule or trip chain describing the sequence of places visited during the simulation. These agenda-like outputs, together with simulated mobility records, can be used to compute aggregate mobility measures such as activity-destination frequencies, origin-destination flows, and trip-count distributions, making the outputs useful for researchers working with origin-destination or check-in-style data as well as higher-frequency trajectory data.






\subsection{Kinematic Trajectory Simulation}\label{subsec:valhalla}
The output of the ABM results in a collection of agendas for all synthetic agents within the simulated world, comprised of instructions for what each agent would like to do over the course of the simulated time horizon assuming no delays due to traffic. The next step is to translate these behavioral decisions into fully-realized kinematic trajectories.

Agenda items encode information about the departure times and modalities of travel for each leg of movement. This information is propagated into the open-source Valhalla routing engine (\url{https://github.com/valhalla/valhalla}) to perform the kinematic fulfillment of the agenda items for all agents. This step uses the OSM road network data by mapping origin and destination buildings to the nearest point on the road network. Additional work is done to ensure continuity of the OSM road network such as  connecting vertices of the road network that have nearly the same location and removing disconnected parts of the network. The resulting fully connected version of the San Francisco OpenStreetMap road network is shared in our repository to enable reproducibility. In addition, Valhalla digests GTFS data to understand public transportation schedules and allow agents to use different types of transportation modes depending on distance between origin and destination and agent ownership of a car or bike.

\section{Dataset Overview and Technical Specifications}
\label{sec:dataset-overview}
%

SF-LIFE is a large-scale simulated movement dataset for the San Francisco Bay Area. It contains complete, noise-free trajectory data for 500{,}000 synthetic agents over a 70-day simulation period, together with activity agendas, agent demographics, building metadata, and supporting network data. The dataset is publicly available at
\url{https://huggingface.co/datasets/sf-life/sf-life}.
The release is organized to support both full-scale experiments and smaller, reproducible workflows: in addition to the full 500{,}000-agent population, the repository provides reduced-scale sub-populations with fewer agents and, where appropriate, coarser temporal sampling rates.

Key characteristics include its massive scale, featuring over three trillion trajectory movement records sampled at 1Hz for each agent over 70 Days of simulation time. The spatial coverage spans the complete San Francisco Bay Area, covering geographic boundaries from a latitude of 37.0°N to 38.5°N and a longitude of 122.0°W to 123.0°W, encompassing a total area of approximately 18,130 square kilometers. The transportation network includes over 40 transit agencies such as BART, Caltrain, AC Transit, and SFMTA, alongside more than 1,000 transit routes across bus and rail services\footnote{We are not able to share GTFS data due to licensing limitations. These can be obtained at \url{https://511.org/open-data/transit}.}. Additionally, it features over 10,000 transit stops and stations across nine counties: Alameda, Contra Costa, Marin, Napa, San Francisco, San Mateo, Santa Clara, Solano, and Sonoma. 


The repository organization is shown in Table~\ref{tab:repository-structure}.
Trajectory files describe agent locations at regular sampled time intervals; agenda files describe the intended activity schedule used by the simulator; demographic files describe agent attributes; the building file provides spatial and semantic information about locations referenced by agents and agendas; and the road network file provides the underlying street network context.

\begin{table}[ht]
\centering
\caption{Repository overview.}
\label{tab:repository-structure}
\small
\begin{tabular}{@{}p{0.27\linewidth}p{0.67\linewidth}@{}}
\toprule
\textbf{Path} & \textbf{Contents} \\
\midrule
\path|data/| & SF-LIFE data.\\
\path|data/<N>_agents/| & Files for a specific sub-population size $N \in \{15,100,1000,10000,500000\}$, including trajectory sub-datasets.\\
\path|data/<N>_agents/metadata/| & Population-specific metadata, including agendas and agent demographics.\\
\path|osm/| & OSM road network data and buildings file.\\
\bottomrule
\end{tabular}
\end{table}

\subsection{Trajectory Sub-datasets}
\label{subsec:trajectory-subdatasets}

The core data product is the set of agent trajectories. Each trajectory record is stored in Parquet format and has the schema shown in Table~\ref{tab:trajectory-schema}. Timestamps are recorded in UTC, coordinates are given as longitude--latitude pairs in WGS84, and the \path|modality| code records the agent's current transportation modality. 

\begin{table}[ht]
\centering
\caption{Trajectory record schema.}
\label{tab:trajectory-schema}
\small
\begin{tabular}{@{}lp{0.20\linewidth}p{0.48\linewidth}@{}}
\toprule
\textbf{Field} & \textbf{Type} & \textbf{Description} \\
\midrule
\path|timestamp| & datetime & UTC timestamp of the sampled trajectory record. \\
\path|agent| & integer & Unique identifier of the synthetic agent. \\
\path|modality| & integer & Transportation modality: 0 = stationary, 1 = walking, 2 = bike, 3 = car, 5 = bus, 6 = rail. \\
\path|longitude| & float & Geographic longitude in decimal degrees. \\
\path|latitude| & float & Geographic latitude in decimal degrees. \\
\bottomrule
\end{tabular}
\end{table}

To reduce the computational burden of working with the full dataset, SF-LIFE provides multiple trajectory sub-datasets that vary along two dimensions: the number of agents and the temporal sampling rate. These sub-datasets are representative subsets of the full simulated population and are intended to support development, debugging, benchmarking, and experiments at different scales. 
Furthermore, the data is provided in two arrangement formats: an agent-centric format (allocating one file per agent) and a bucketed format. 
The by-agent layout stores one trajectory file per agent, which is convenient for small and medium-sized subsets. The bucketed layout groups multiple agents into shared Parquet files to make the larger 10{,}000-agent and 500{,}000-agent releases more manageable. For bucketed releases, the corresponding agent-to-bucket mapping file identifies the bucket that contains a given agent's trajectory.
Table~\ref{tab:trajectory-subdatasets} lists the available combinations, their storage layouts, and compressed volume sizes.


\begin{table}[htbp]
\centering
\caption{Trajectory sub-datasets.}
\label{tab:trajectory-subdatasets}

\small
\begin{tabular}{@{}l@{\hspace{0.1em}}ccccc@{}}
\toprule
\multirow{2}{*}{\begin{tabular}{@{}c@{}}\textbf{Temporal}\\\textbf{Sampling Rate}\end{tabular}}
& \multicolumn{5}{c}{\textbf{Number of Agents ($N$)}} \\ 
\cmidrule(l){2-6}
& \textbf{15} & \textbf{100} & \textbf{1,000} & \textbf{10,000} & \textbf{500,000} \\ 
\midrule

\textbf{1s}         & $\bullet$ 0.24G   & $\times$    & $\times$      & $\times$       & $\blacksquare$ 443G       \\
\textbf{5s}         & $\times$    & $\bullet$ 0.36G   & $\bullet$ 3.60G     & $\blacksquare$ 37.8G      & $\times$        \\
\textbf{1min}       & $\times$   & $\bullet$ 0.06G    & $\bullet$ 0.67G     & $\blacksquare$ 2.70G      & $\blacksquare$ 6.20G      \\
\textbf{10min}      & $\times$   & $\times$    & $\bullet$  0.07G    & $\blacksquare$ 0.08G      & $\blacksquare$ 1.70G       \\
\textbf{30min}      & $\times$   & $\times$    & $\bullet$  0.02G    & $\blacksquare$ 0.02G      & $\blacksquare$ 0.78G       \\
\textbf{60min}      & $\times$   & $\times$    & $\bullet$ 0.01G     & $\blacksquare$ 0.01G        & $\blacksquare$  0.57G     \\ \bottomrule

\multicolumn{6}{@{}l}{\rule{0pt}{3ex}\small \textbf{Legend:} $\bullet$ = agent-centric format; $\blacksquare$ = bucketed format} \\

\end{tabular}
\end{table}

\subsection{Structured Agenda and Activity Records}
\label{subsec:agenda-records}

In addition to sampled trajectories, each population directory contains an agenda file named \path|sf-life_agenda_<N>_agents.parquet|. These files describe the intended activity schedule used by the simulation. They should be interpreted as structured agenda and activity records rather than as a replacement for the observed trajectories: agents attempt to follow their agendas, but realized movement may differ from the intended timestamps or destinations because of travel-time constraints and simulation dynamics.

Each agenda record includes an agent, an intended timestamp, a referenced building, an activity type, and an intended transport mode. 
The main fields are summarized in Table~\ref{tab:agenda-schema}.
If desired, these agendas can separately be used as origin-destination (OD) data, if this format of segmented trip records is preferred.

 If the \path|activity_type| is \path|AtPudos|, the agent is either at a pick-up or drop-off for another agent. Records with \path|activity_type = Transport| encode planned movement between activities. Public-transit trips appear in the agenda as \path|multimodal| because an agent typically combines public transit with access and egress travel, such as walking to and from stations or stops. The realized movement mode at any sampled trajectory timestamp is given separately by the trajectory \path|modality| field.

\begin{table}[h]
\centering
\caption{Agenda record schema.}
\label{tab:agenda-schema}
\small
\begin{tabular}{@{}lp{0.66\linewidth}@{}}
\toprule
\textbf{Field} & \textbf{Description} \\
\midrule
\path|timestamp| & Intended arrival time at a location or intended start time of a trip. \\
\path|agent| & Agent identifier. \\
\path|building| & Building identifier referenced by the agenda entry. For transport entries, this refers to the origin building. \\
\path|activity_type| & Intended activity, from  \path{AtHome}, \path|AtWork|, \path|AtSchool|, \path|AtRestaurant|, \path|AtRecreation|, \path{AtWorship}, \path|Transport|, \path{AtErrand}, or \path{AtPudos}. \\
\path|transport_mode| & Intended mode for transport entries, with values such as \path|pedestrian|, \path|bicycle|, \path|auto|, and \path|multimodal|; this value is \path|none| for non-transport activities. \\
\bottomrule
\end{tabular}
\end{table}

\subsection{Demographics and Building Metadata}
\label{subsec:demographics-buildings}

The dataset includes two main forms of contextual metadata. First, each population directory contains a demographics file named \path|sf-life_agent_demographics_<N>_agents.csv|. For the full population, this file contains 500{,}000 rows; for reduced-scale releases, it identifies the agents included in that sub-population. Each record contains the agent identifier, age, gender, home building, and agent type. The \path|home_building| field references the global building table, while \path|agent_type| describes the agent's broad behavioral role, such as student, worker, or homemaker.

Second, the global building file \path|data/metadata/sf-life_building_mapping.csv| defines the locations used by the simulation. It contains a \path|buildingId| primary key, geographic coordinates, and a semantic category, one of residential, workplace, religion, education, restaurant, or recreation. The building identifiers are referenced by the demographics files through \path|home_building| and by the agenda files through \path|building|. This relational structure allows researchers to connect agent attributes, intended activities, and realized movements through shared identifiers.

\begin{figure}[hbpt]
    \centering
    \includegraphics[width=0.9\linewidth]{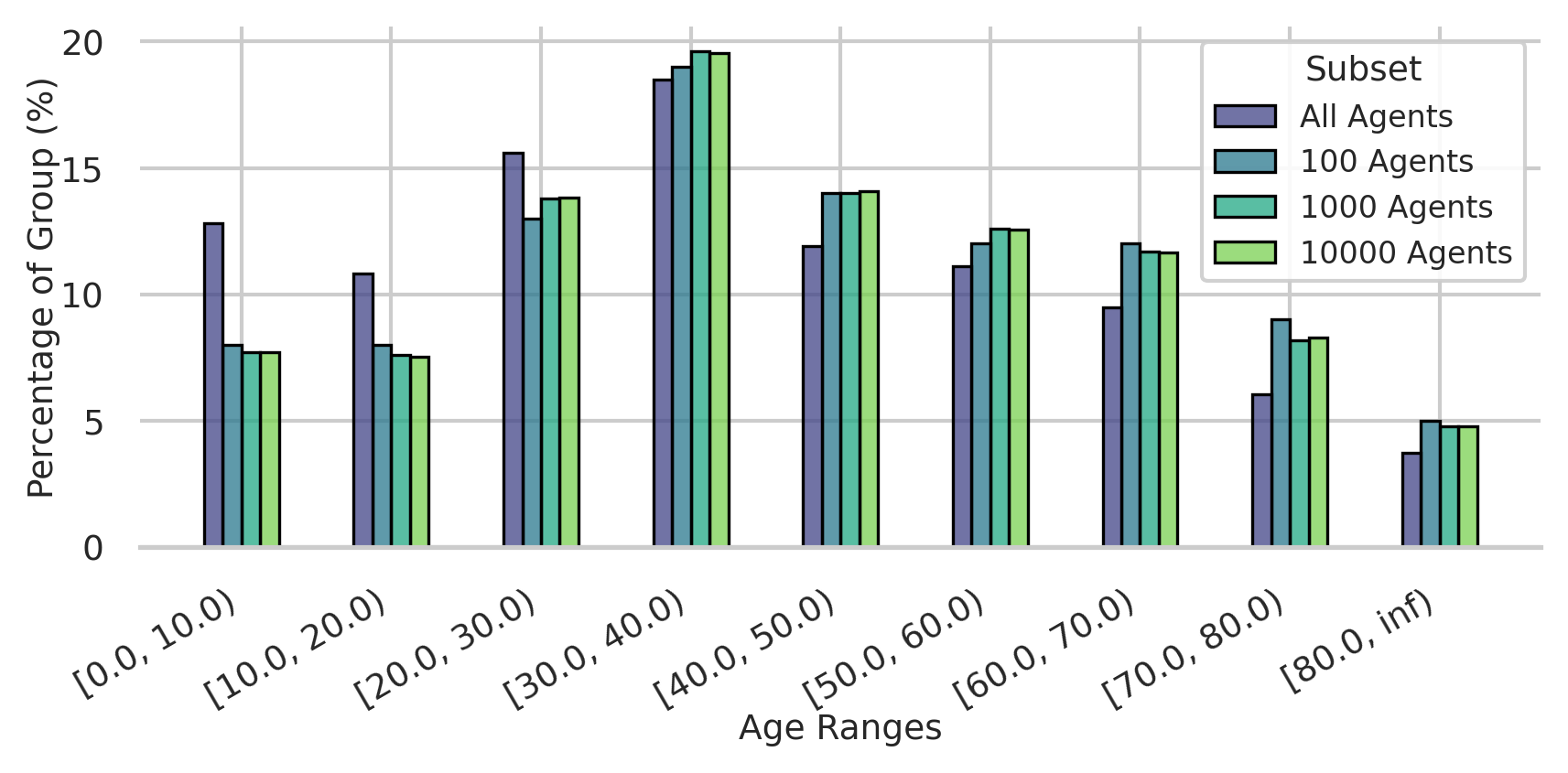}
    \caption{Agent age distribution across subsets.}
    \label{fig:age_plot}
\end{figure}

The synthetic population in SF-LIFE is designed to reflect a realistic demographic structure while maintaining statistical consistency across sampled subsets. As shown in Figure~\ref{fig:age_plot}, the age distribution is well-balanced across major cohorts, with the largest proportion of agents falling within the 30–49 age range, followed by younger (0–17) and early working-age (18–29) groups. Older populations (50–64 and 65+) are also represented at meaningful levels, ensuring that the dataset captures the full lifecycle of mobility behaviors. This distribution aligns with expected urban demographics, where working-age individuals dominate overall activity levels, while younger and older populations contribute distinct travel patterns, such as school-related trips and reduced mobility frequency.

\begin{figure}[h]
    \centering
    \includegraphics[width=\linewidth]{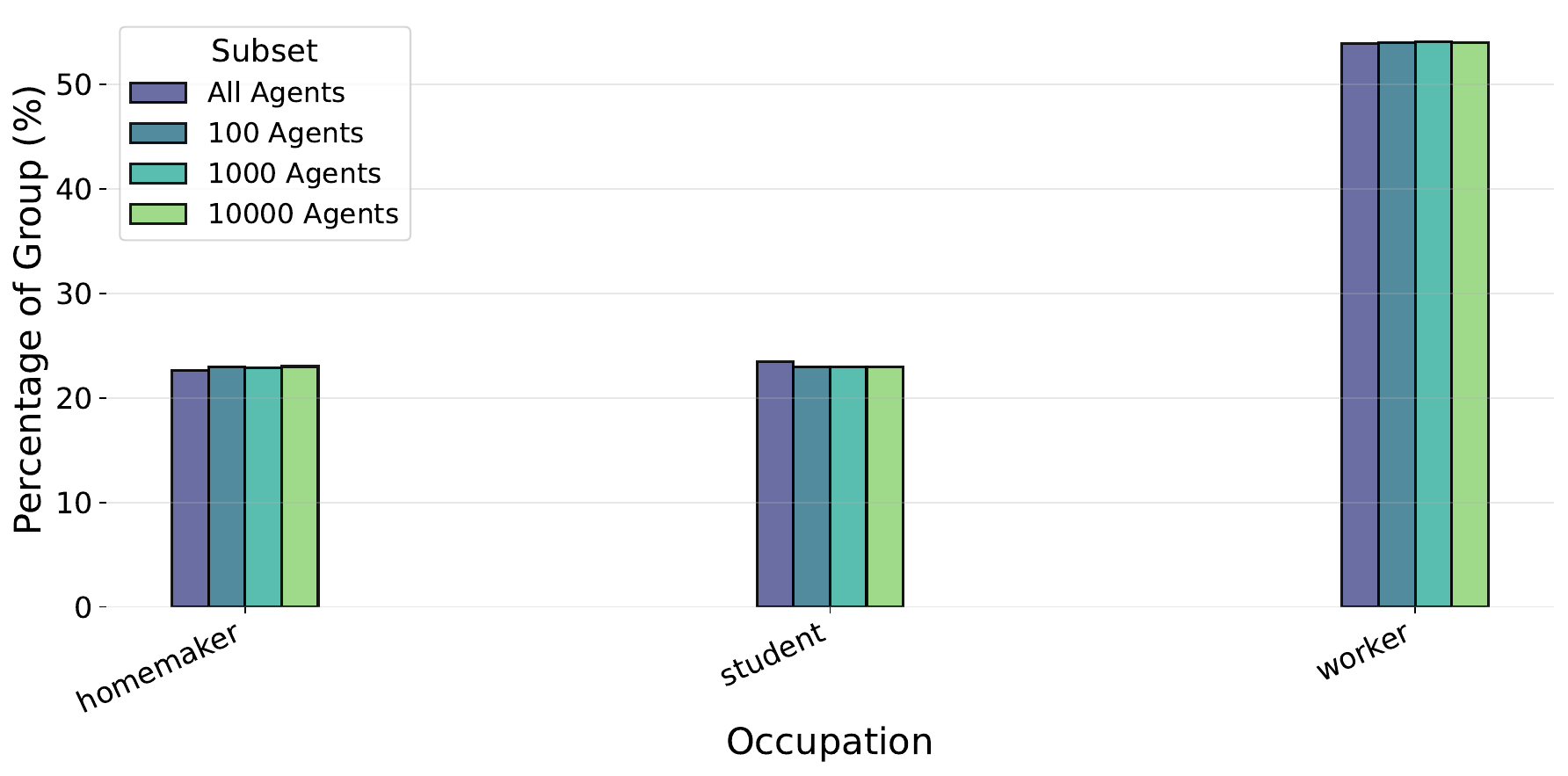}
    \caption{Agent occupations across subsets.}
    \label{fig:agent_type}
\end{figure}

\begin{figure}[h]
    \centering
    \includegraphics[width=\linewidth]{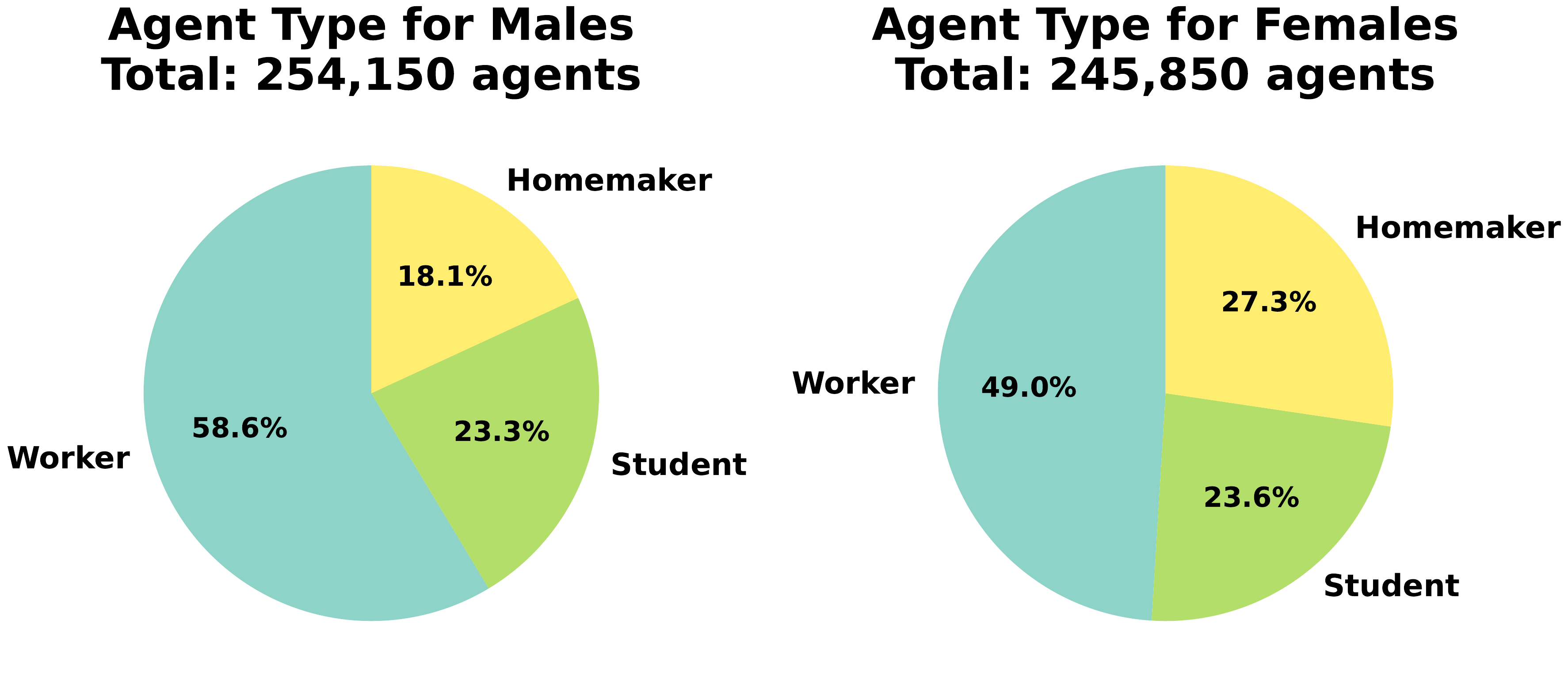}
    \caption{Gender vs agent type.}
    \label{fig:gender-type}
\end{figure}

Agent roles further reinforce behavioral realism. As illustrated in Figure~\ref{fig:agent_type}, the population is composed of workers, students, and homemakers, with workers forming the majority group. This composition directly drives the simulation’s activity patterns, as workers and students are associated with mandatory daily trips, while homemakers exhibit more flexible, needs-driven mobility. The gender-disaggregated breakdown in Figure~\ref{fig:gender-type} highlights subtle but important differences: male agents are more likely to be classified as workers, whereas female agents have a higher proportion of homemaker roles, with student representation remaining relatively consistent across genders. These distinctions introduce heterogeneity in daily schedules and trip purposes, which is critical for generating realistic aggregate mobility patterns.

\begin{figure}[h]
    \centering
    \includegraphics[width=0.5\linewidth]{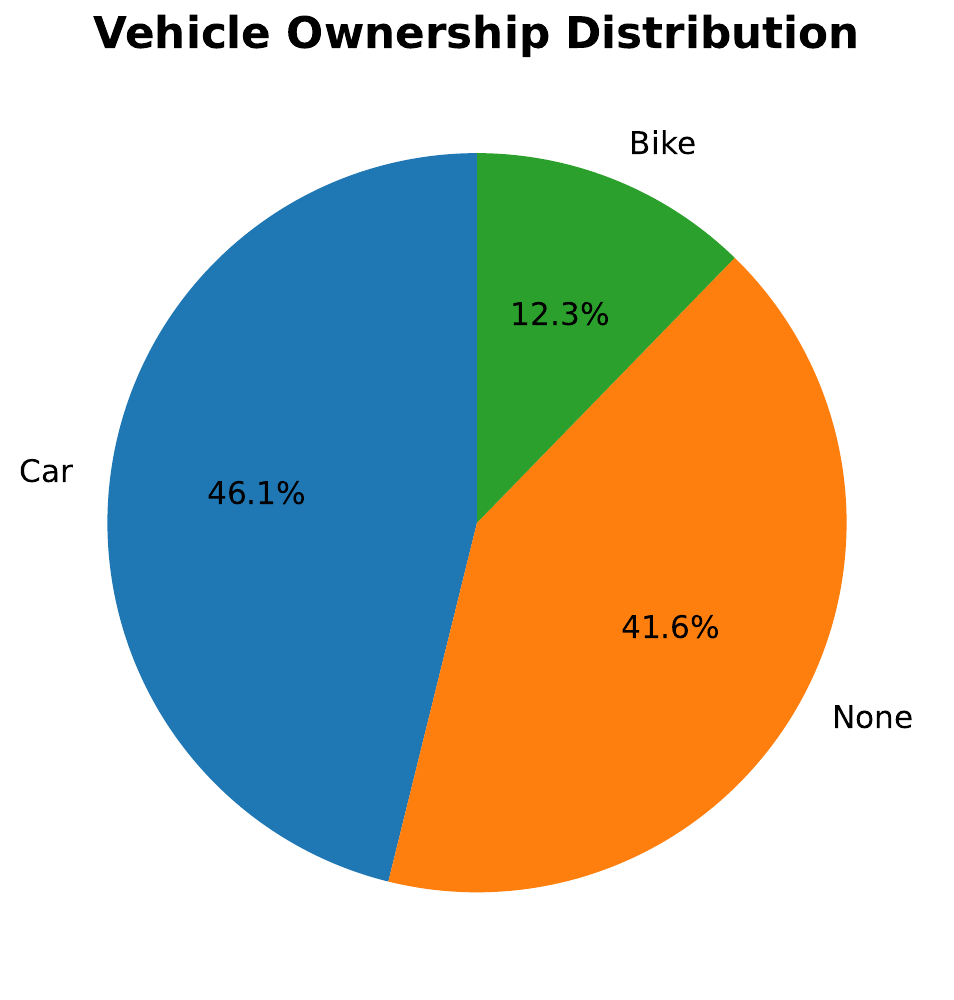}
    \caption{Vehicle Ownership.}
    \label{fig:vehicle-owner}
\end{figure}

Finally, Figure~\ref{fig:vehicle-owner} presents the distribution of vehicle ownership, a key determinant of transportation mode choice. Approximately half of the population has access to a car, while a substantial fraction relies on non-car modes or has no private vehicle access, with a smaller segment using bicycles. This balance ensures meaningful interaction between private and public transportation systems within the simulation. By jointly modeling age, occupation, gender, and vehicle ownership, SF-LIFE captures the primary demographic drivers of mobility behavior, enabling a more realistic analysis of travel demand, modal choice, and accessibility across different population segments.

\subsection{Road and Public-Transit Reference Data}
\label{subsec:network-reference-data}

Openstreetmap data for the San Francisco Bay area is included as the road-network file \path|osm/roads.osm|, which provides the street-network context for the simulated movements. 
%
GTFS schedule, route, or stop files are not included, but users who need the public-transit network and schedules can obtain current Bay Area GTFS and GTFS-Realtime feeds from the 511 SF Bay Open Data Transit portal
\url{https://511.org/open-data/transit}.

\section{Qualitative Analysis}
\begin{figure}[h]
    \centering
    \includegraphics[width=.9\linewidth]{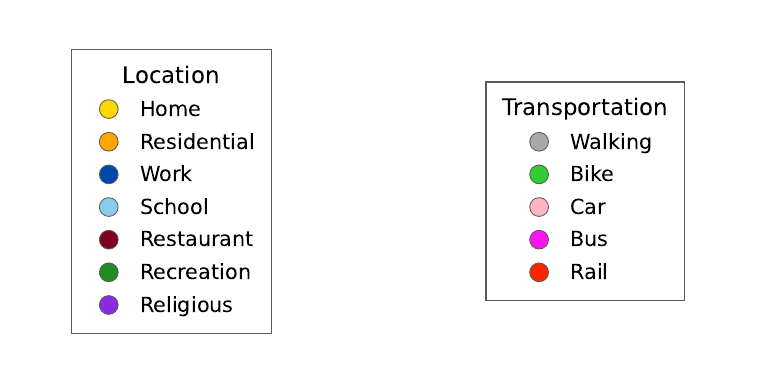}
    \caption{List of colors and their corresponding location or transportation type.}
    \label{fig:legend}
\end{figure}

We demonstrate the realism of our agent simulations by examining their movement patterns using calendar plots and trajectory maps. For ease of interpretation, all spatial data and transportation methods follow the color-coding scheme detailed in Figure~\ref{fig:legend}. While the trajectory maps emphasize movement and location at the expense of temporal specifics, the calendar plots prioritize the pattern, duration, and nature of activities while omitting spatial details. Taken together, these complementary visualizations provide a comprehensive qualitative analysis of the simulated environment.

\subsection{Calendar Plots}
We illustrate the behavioral patterns of selected agents through calendar plots, which effectively highlight the semantics of trajectory data. Because habits are a fundamental driver of human mobility, they are clearly projected within these temporal visualizations~\cite{sereshgi_calendar2025}.

\begin{figure}[p]
    \centering
    \includegraphics[width=\linewidth]{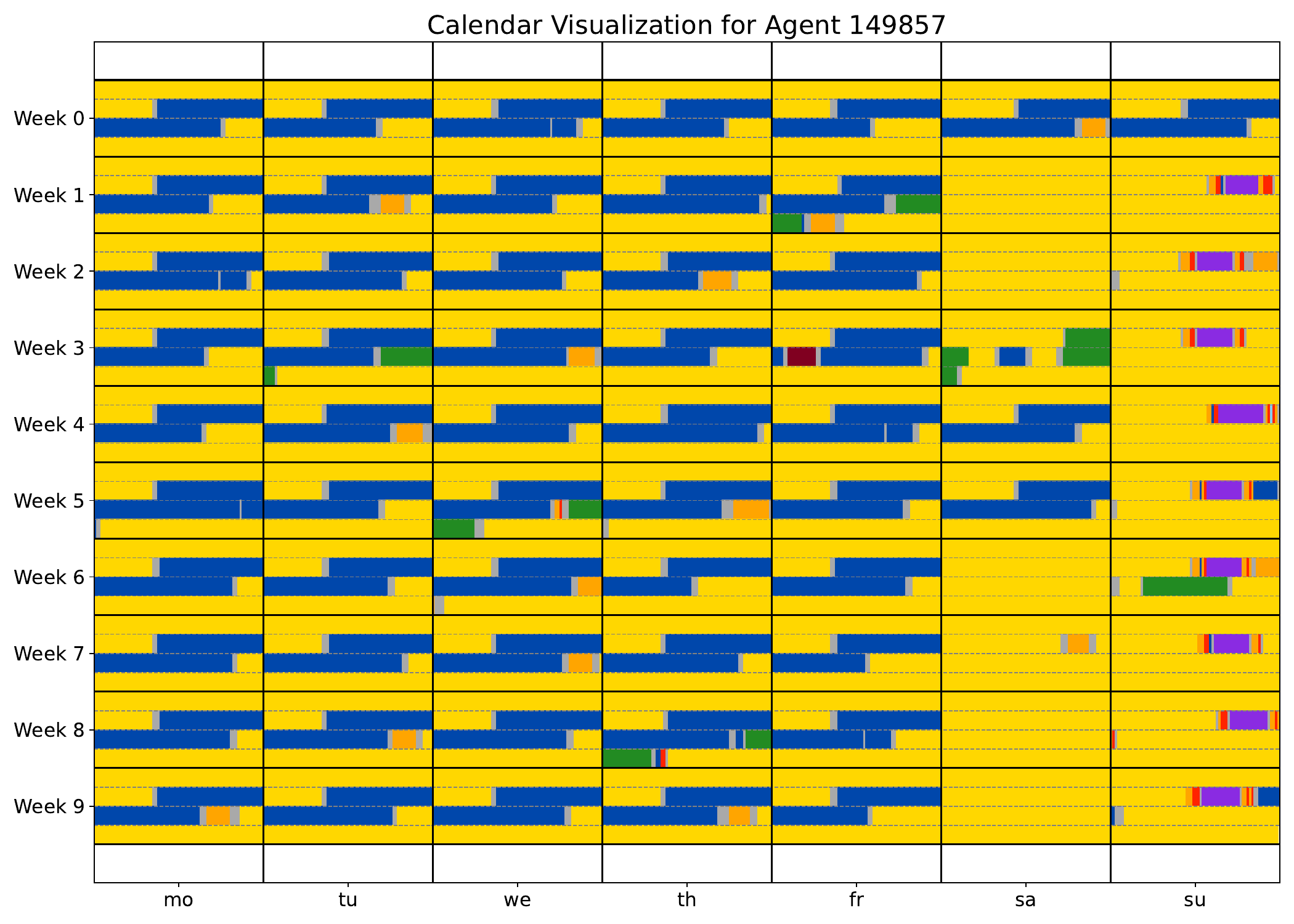}
    \caption{Life patterns of agent $149857$ (worker).}
    \label{fig:cal149857}
\end{figure}

\begin{figure}[p]
    \centering
    \includegraphics[width=\linewidth]{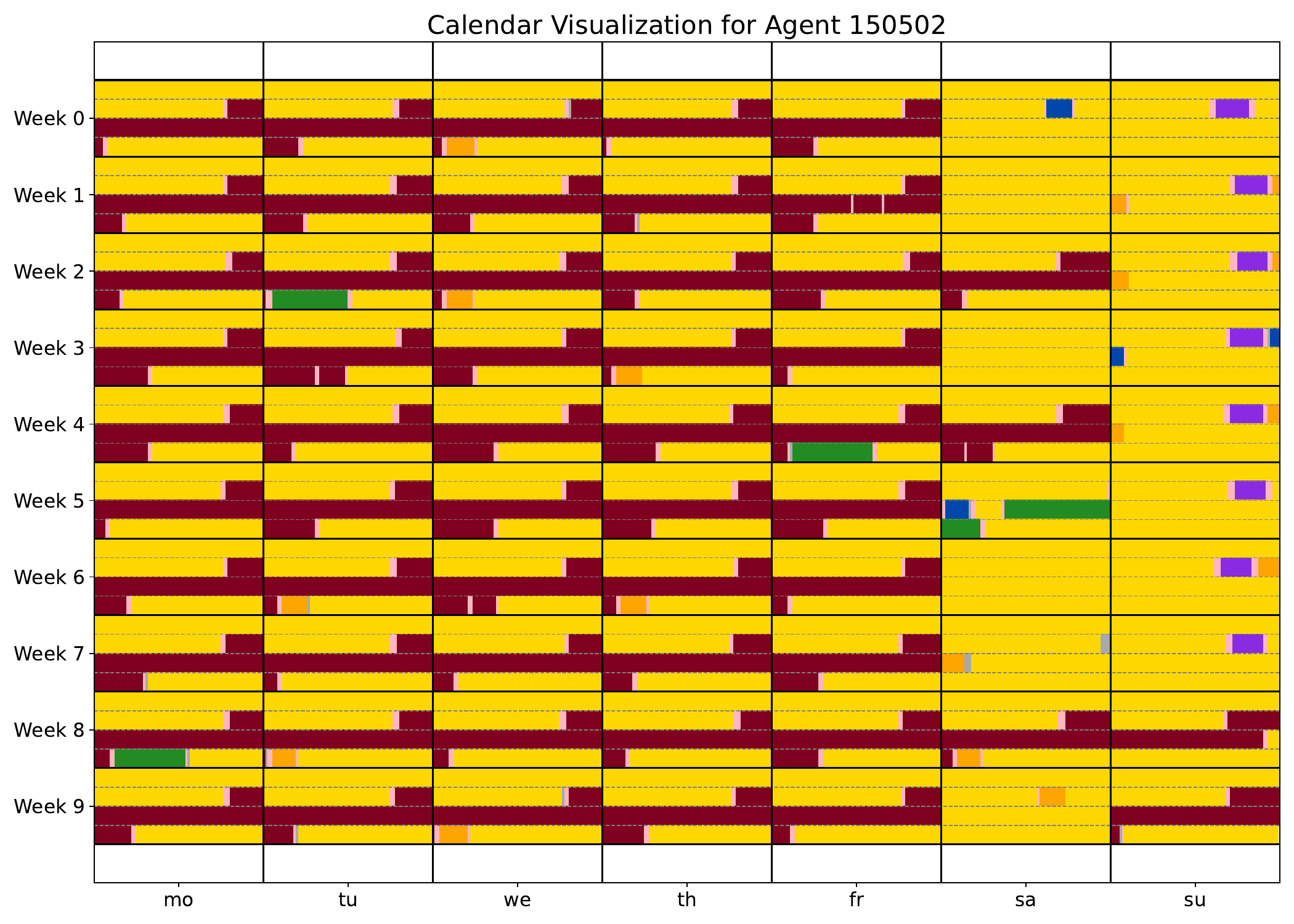}
    \caption{Life patterns of agent $150502$ (worker).}
    \label{fig:cal150502}
\end{figure}

As shown in Figure~\ref{fig:cal149857}, Agent 149857 exhibits a structured weekday routine, commuting to work on foot and occasionally visiting a gym in the evenings. On weekends, this agent typically attends church, utilizing rail transport. In contrast, Figure~\ref{fig:cal150502} depicts an agent employed at a restaurant. This individual commutes primarily by car and occasionally works weekend shifts, though they generally maintain a weekend church visit. Notably, this agent avoids public transportation entirely.

\begin{figure}[p]
    \centering
    \includegraphics[width=\linewidth]{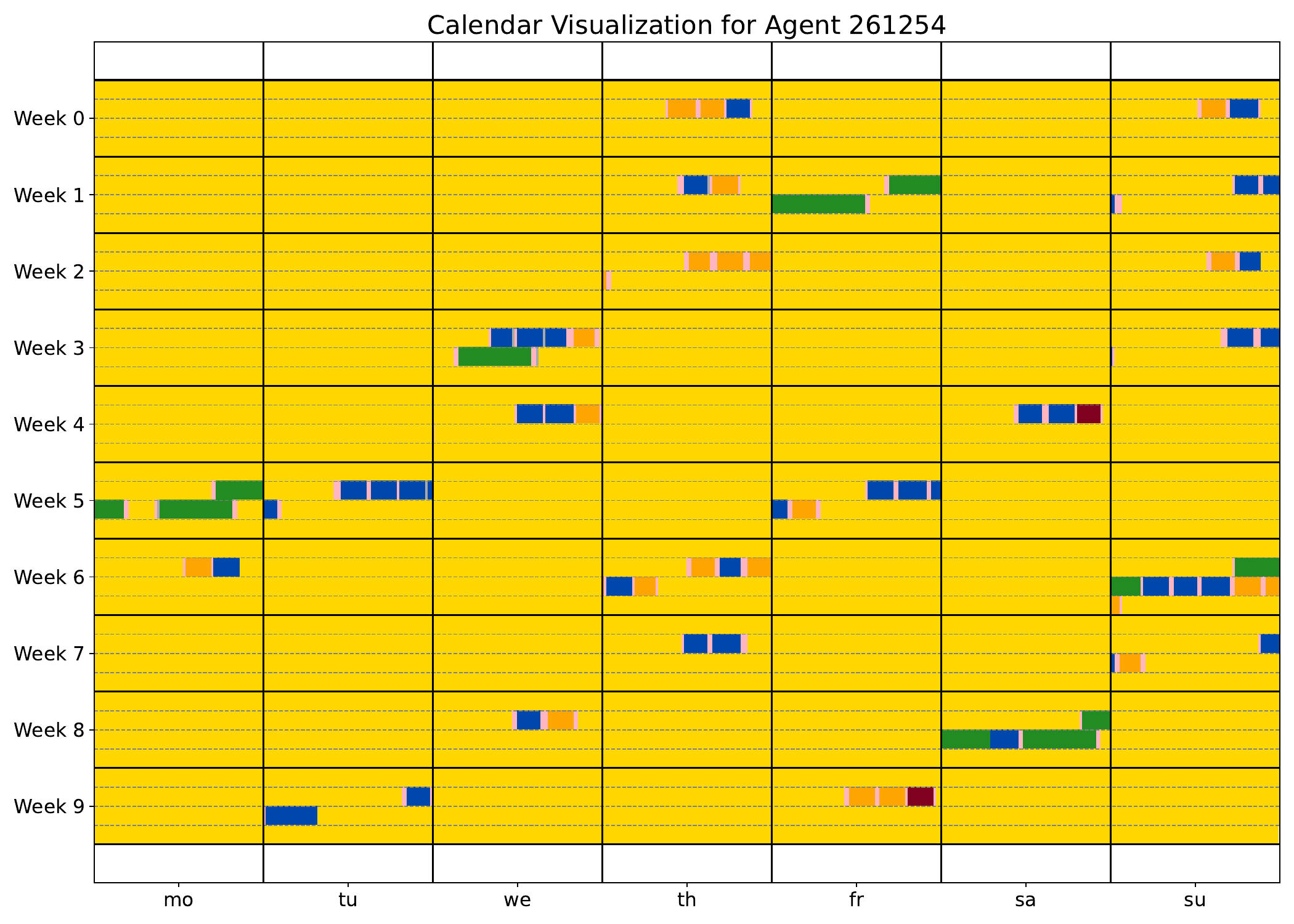}
    \caption{Life patterns of agent $261254$ (homemaker).}
    \label{fig:cal261254}
\end{figure}

\begin{figure}[p]
    \centering
    \includegraphics[width=\linewidth]{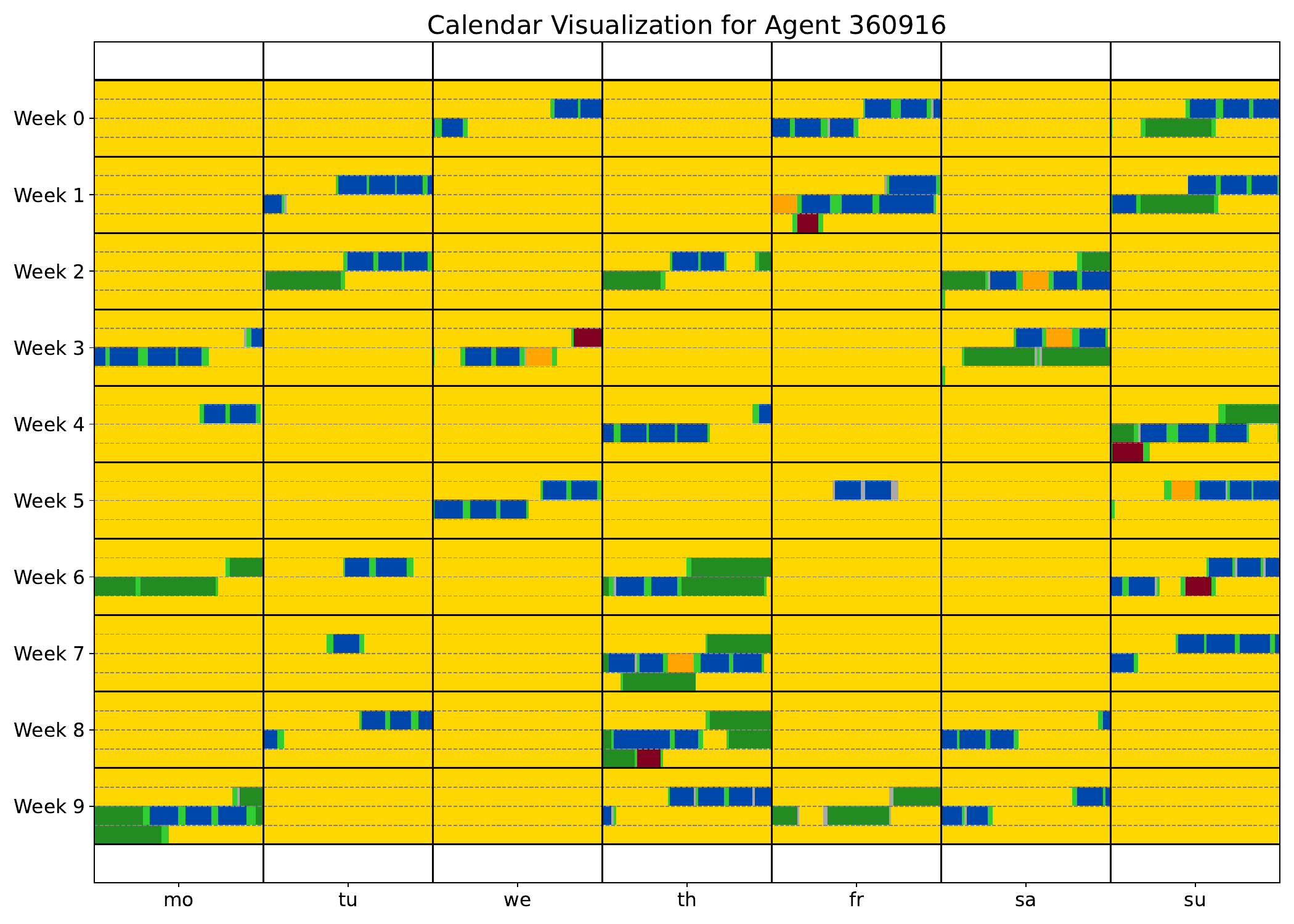}
    \caption{Life patterns of agent $360916$ (homemaker).}
    \label{fig:cal360916}
\end{figure}

The mobility of a "homemaker" agent, illustrated in Figure~\ref{fig:cal261254}, reveals a more flexible schedule without attending work or school mainly using walking as their transportation mode to perform errands. While broad trends are less rigid, local patterns remain discernible; for instance, this agent typically visits multiple locations per outing. Similarly, Figure~\ref{fig:cal360916} captures another homemaker agent who relies predominantly on a bicycle for their errands and spends more time on recreational activities.

\begin{figure}[p]
    \centering
    \includegraphics[width=\linewidth]{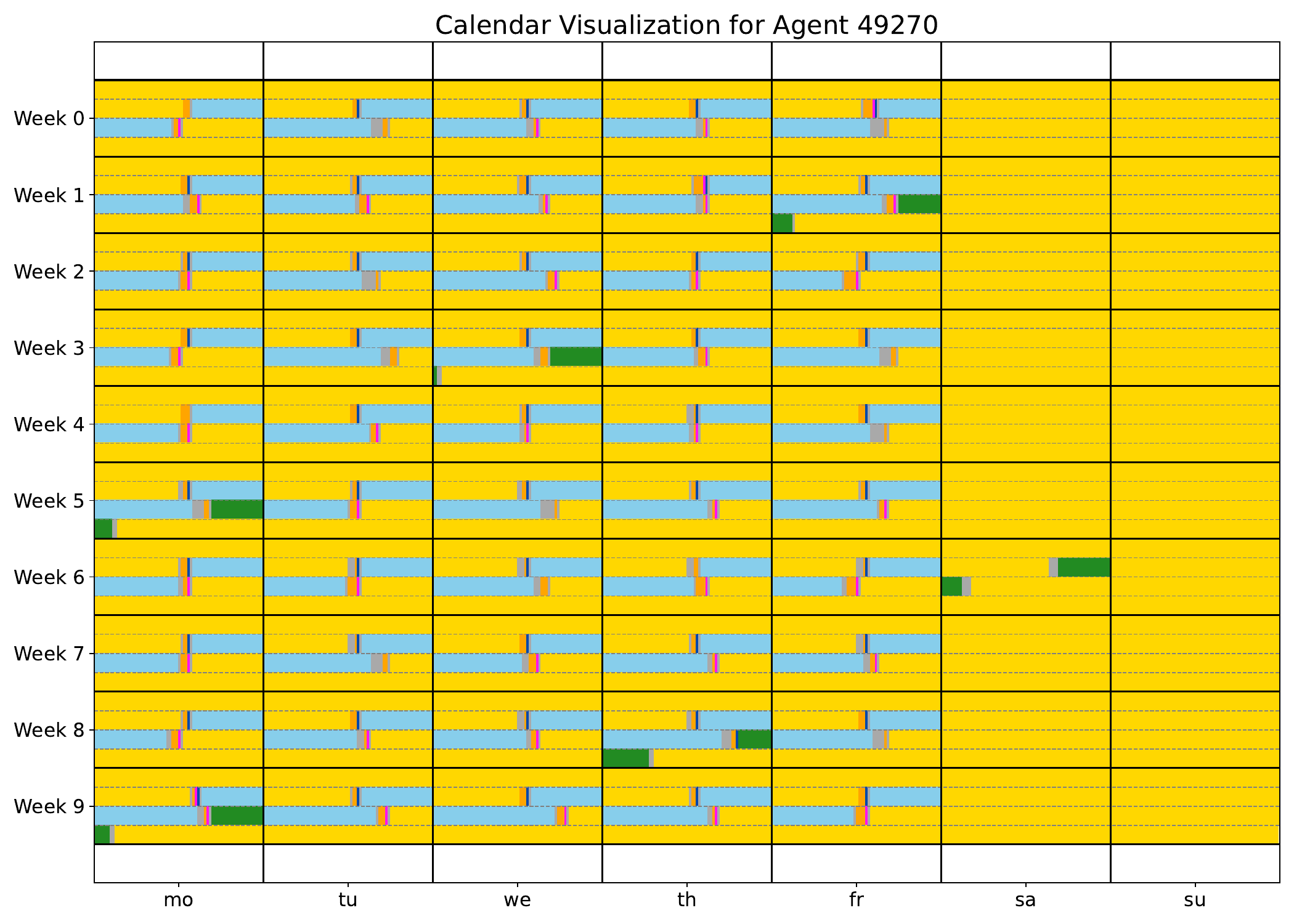}
    \caption{Life patterns of agent $49270$ (student).}
    \label{fig:cal49270}
\end{figure}

\begin{figure}[p]
    \centering
    \includegraphics[width=\linewidth]{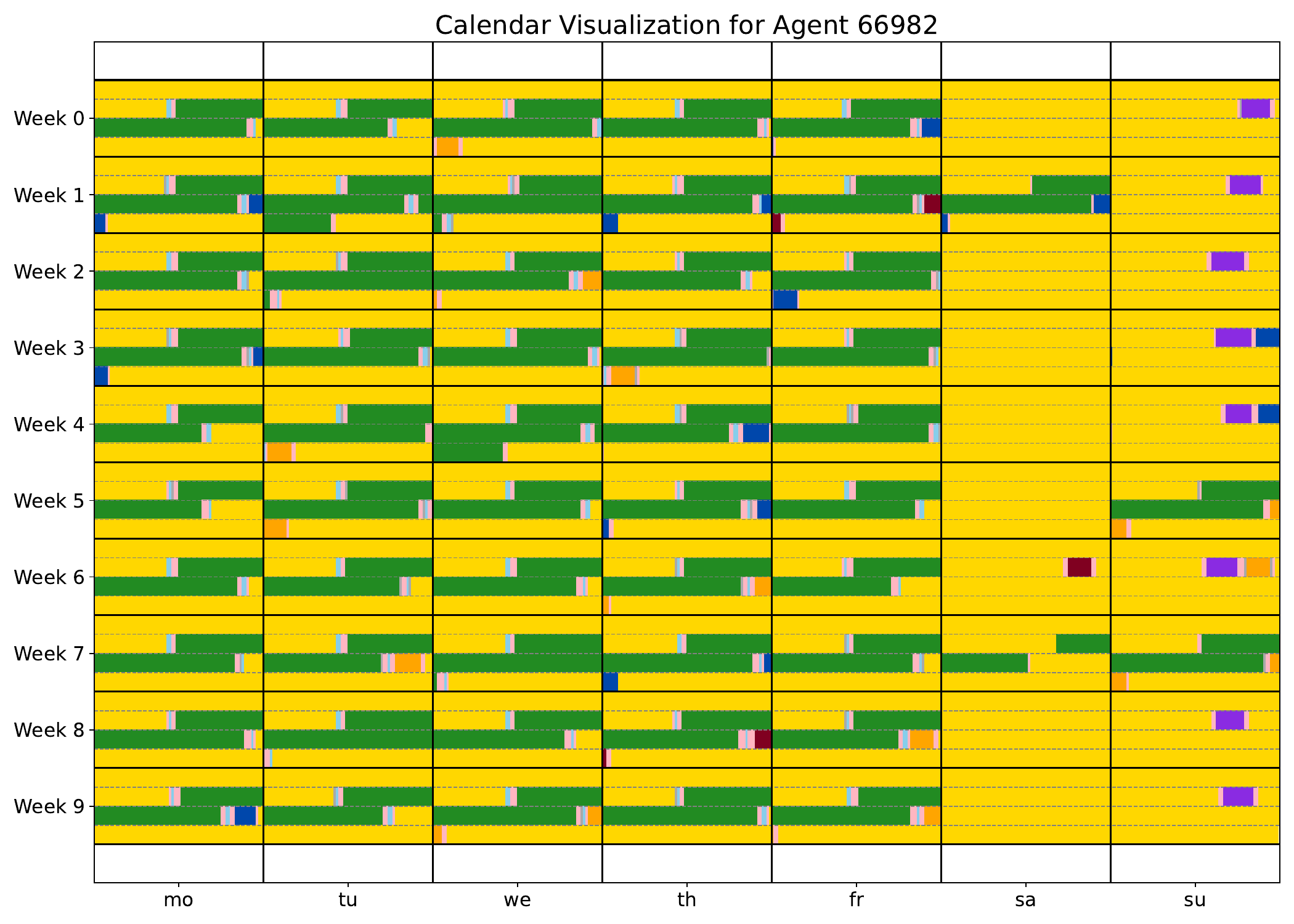}
    \caption{Life patterns of agent $66982$ (worker).}
    \label{fig:cal66982}
\end{figure}

Educational routines are also captured, as seen in the student profile in Figure~\ref{fig:cal49270}. This agent attends school daily via bus or on foot, occasionally visiting recreational sites after classes, while remaining largely at home on weekends. Agent 66982 (Figure~\ref{fig:cal66982}) works at a recreational site. They transition between driving and walking and frequently socialize at the homes of their friends or family immediately after their shifts.

\begin{figure}[t]
    \centering
    \includegraphics[width=\linewidth]{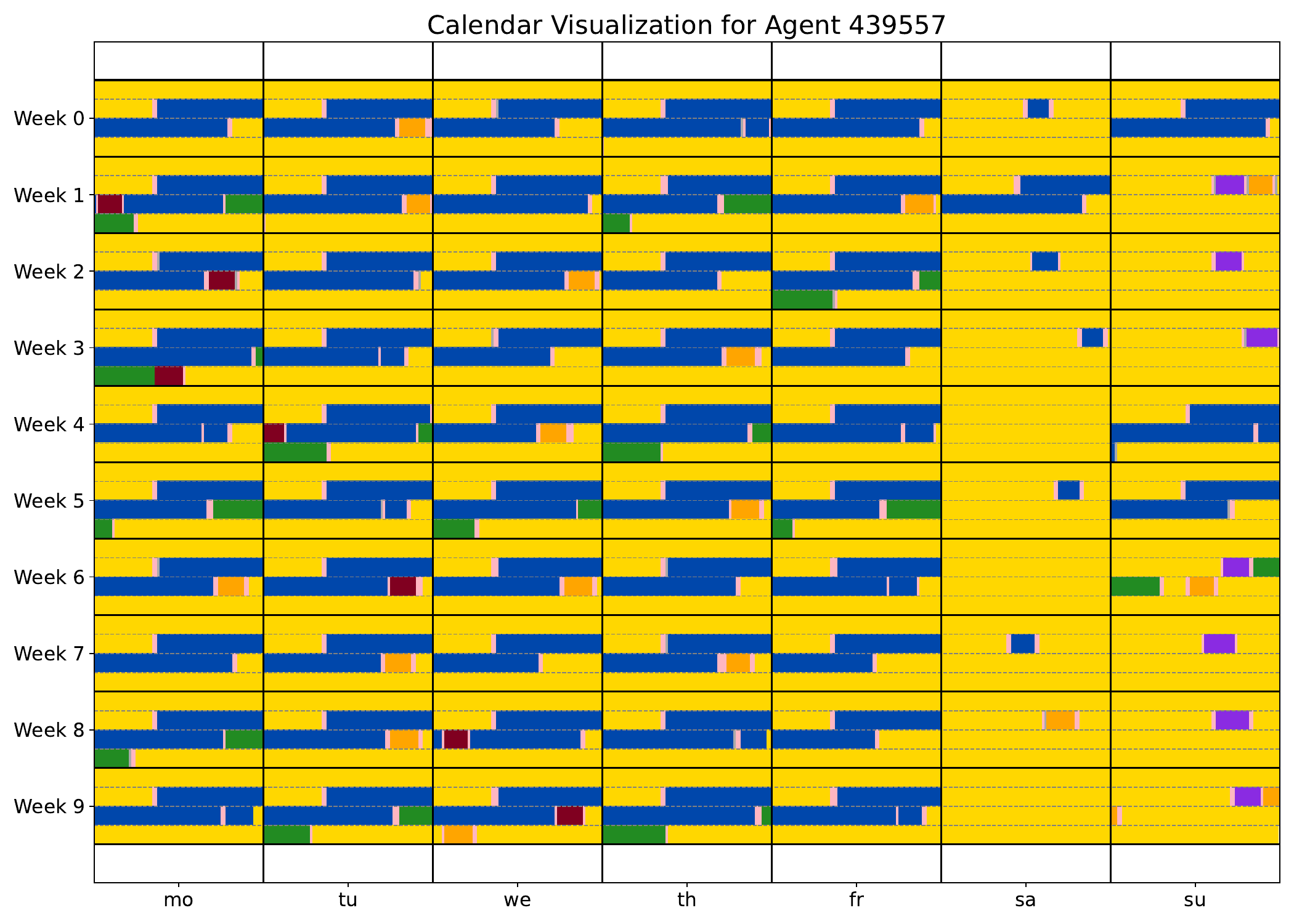}
    \caption{Life patterns of agent $439557$ (worker).}
    \label{fig:cal439557}
\end{figure}

\begin{figure}[t]
    \centering
    \includegraphics[width=\linewidth]{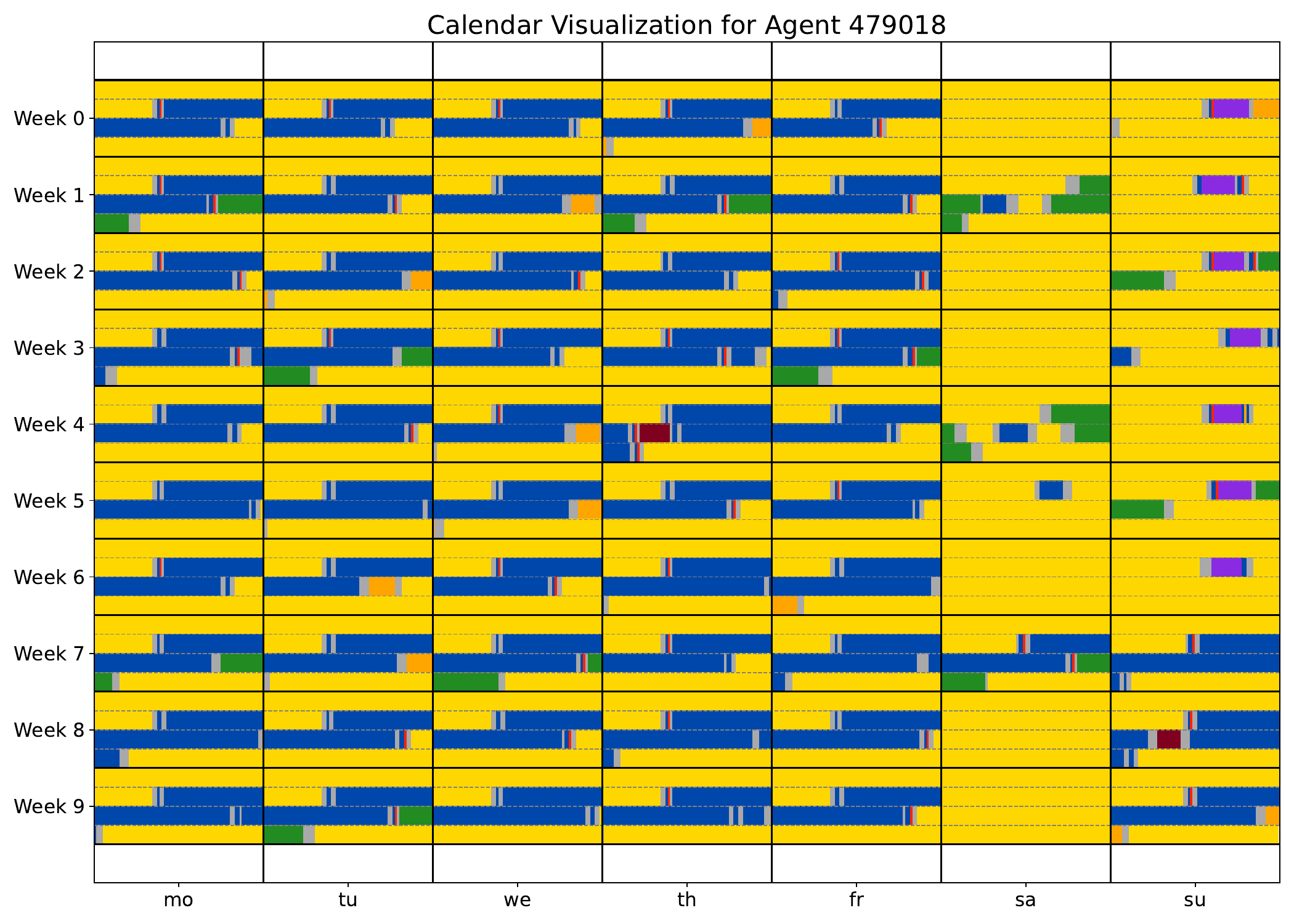}
    \caption{Life patterns of agent $479018$ (worker).}
    \label{fig:cal479018}
\end{figure}

Finally, Figures~\ref{fig:cal439557} and \ref{fig:cal479018} present agents with superficially similar schedules but diverging lifestyle choices. Agent 439557 dines at a restaurant weekly and utilizes a personal vehicle for commuting. In contrast, agent 479018 rarely visits restaurants and prefers walking or taking the train, highlighting how individual preferences differentiate agents with otherwise similar temporal constraints.

\subsection{Trajectories}
\begin{figure}[b]
\vspace*{-1ex}
    \centering
    \includegraphics[width=0.9\linewidth]{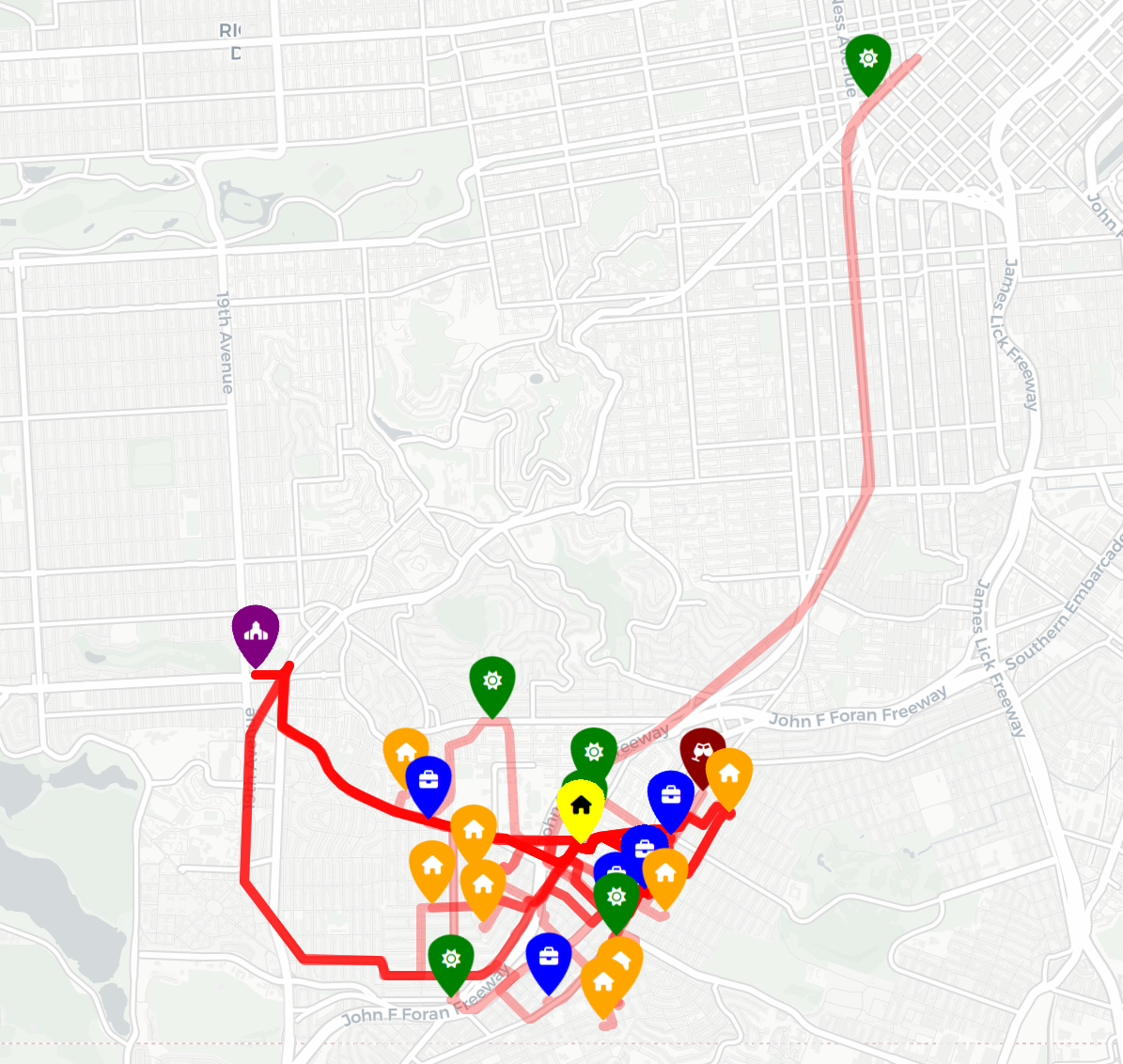}
    \caption{Agent 149857 (worker). Staypoint colors indicate location type and follow calendar plot key.}
    \label{fig:map149857}
\end{figure}

\begin{figure}[b]
    \centering
    \includegraphics[width=0.9\linewidth]{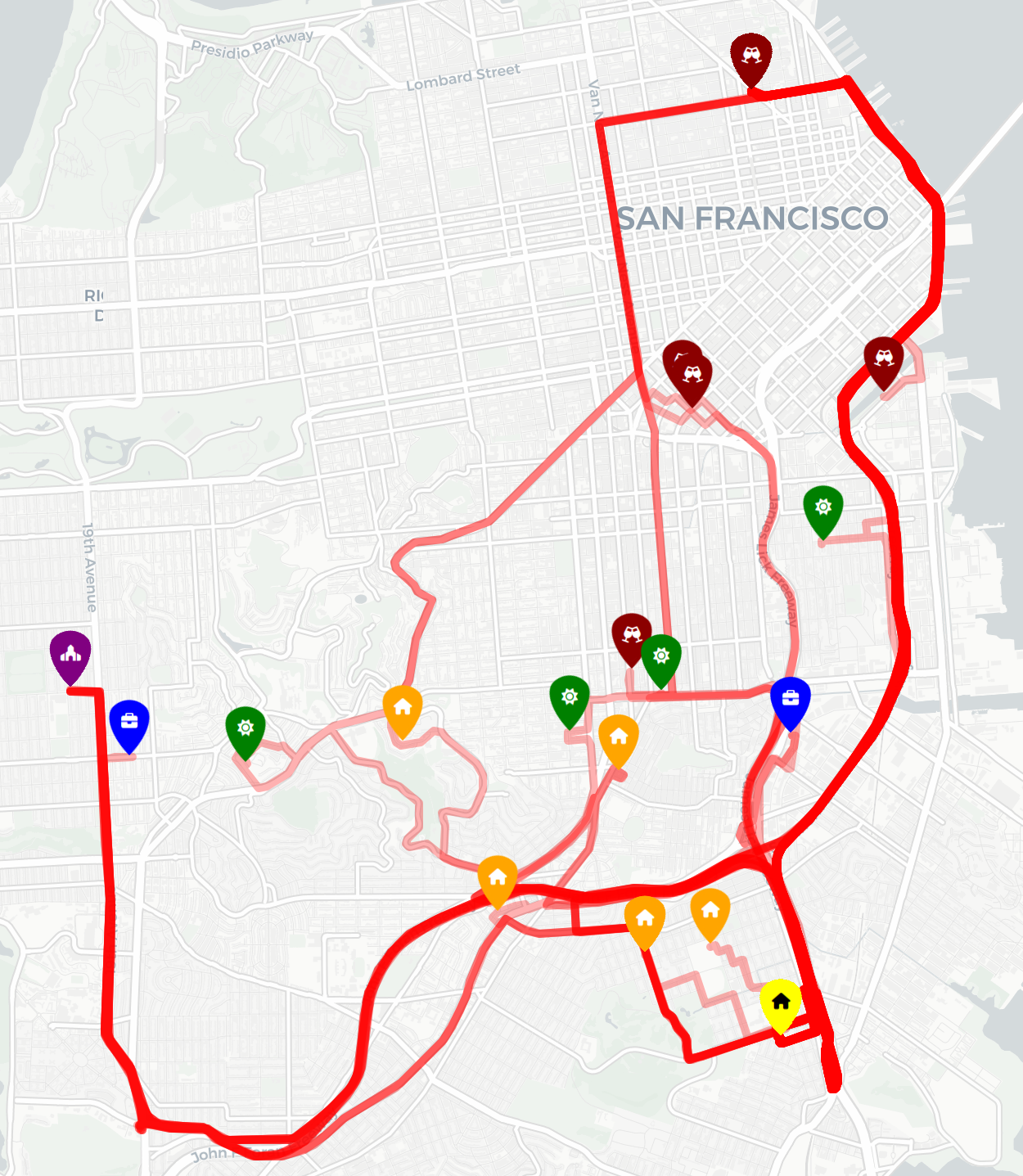}
    \caption{Agent 150502 (worker). Staypoint colors indicate location type and follow calendar plot key.}
    \label{fig:map150502}
\end{figure}

We include visuals of the trajectories of selected agents to display their behavior and patterns. For instance, in Figure~\ref{fig:map149857}, we see the trajectory of agent 149857, who spends most of the simulation period visiting residential, workplace, and recreational locations around their home (yellow). However, we also see a single trip to a distant recreational location which deviates from their norm, as well as regular trips to a single religious location.
Agent 150502 Figure~\ref{fig:map150502} travels further from their residence on average. This agent regularly visits restaurants, and from the calendar plot in Figure~\ref{fig:cal150502}
we see they likely work in restaurants. Just like agent 149857, we

\noindent
 see regular trips to a religious location, and occasional trips to residential and recreational locations.

\begin{figure}[b]
    \centering
    \includegraphics[width=0.9\linewidth]{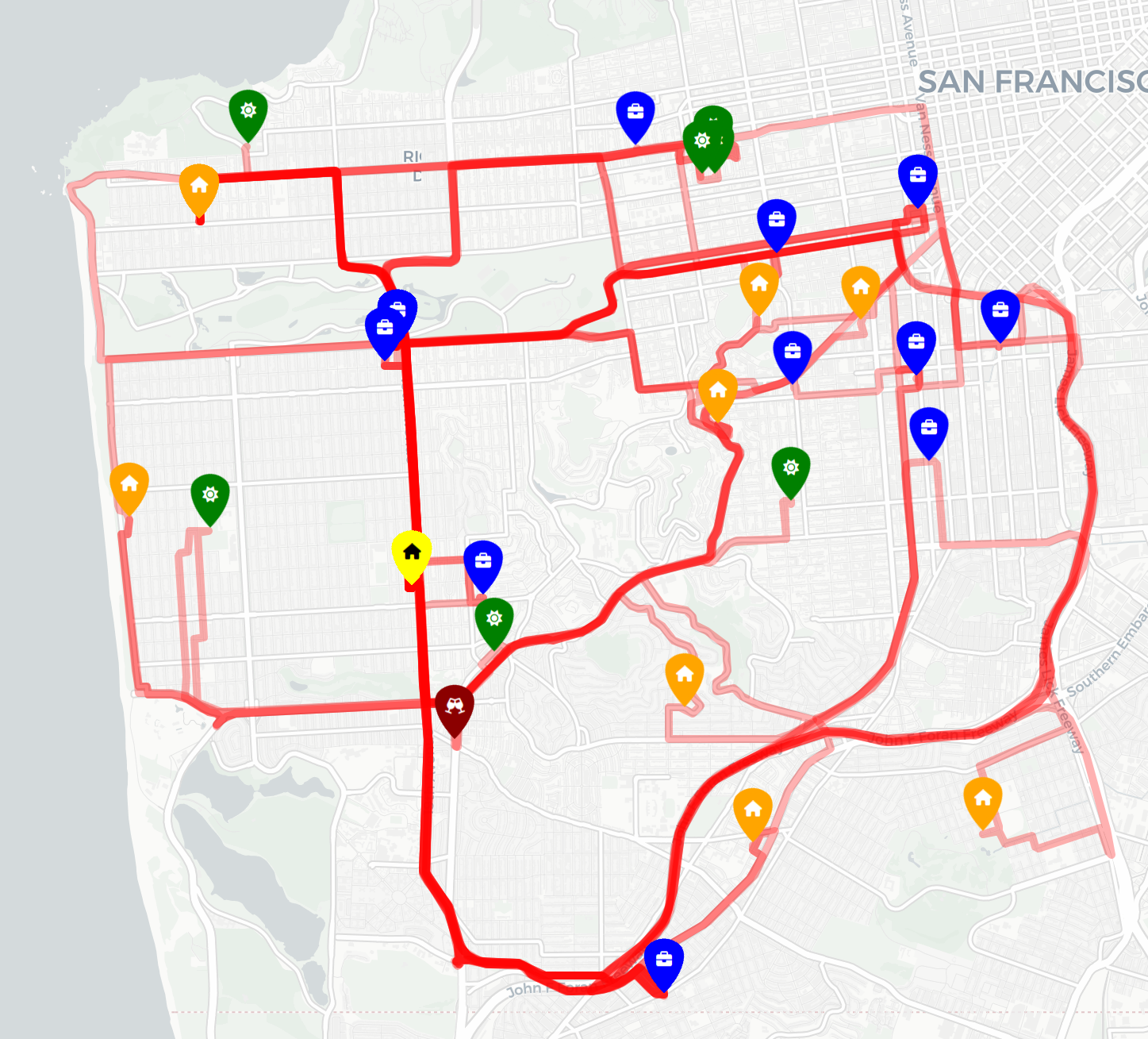}
    \caption{Agent 261254 (homemaker). Staypoint colors indicate location type and follow calendar plot key.}
    \label{fig:map261254}
\end{figure}

\begin{figure}[b]
    \centering
    \includegraphics[width=0.9\linewidth]{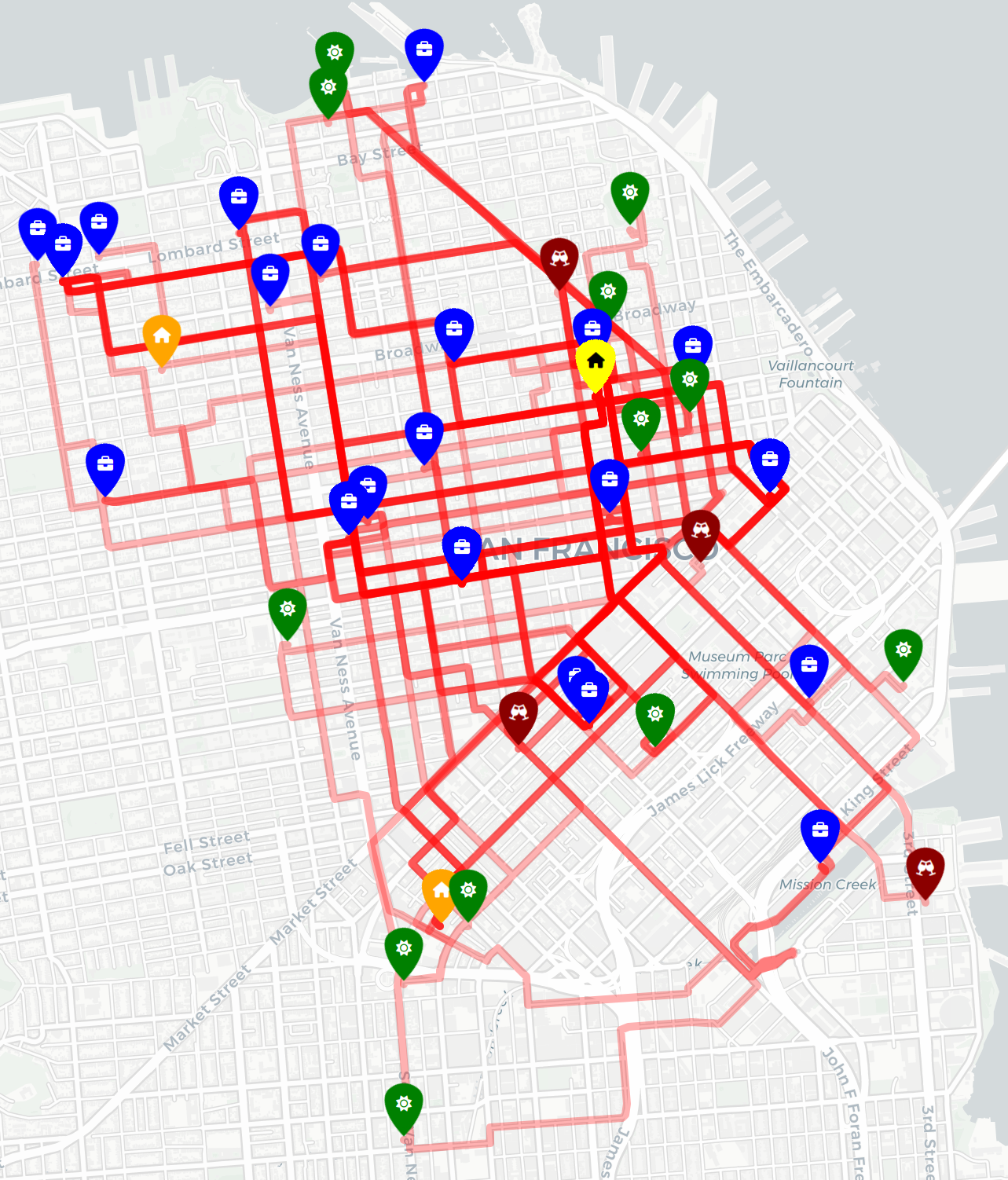}
    \caption{Agent 360916 (homemaker). Staypoint colors indicate location type and follow calendar plot key.}
    \label{fig:map360916}
\end{figure}

Figure~\ref{fig:map261254} shows agent 261254, who does not appear to have many regular locations, and may also travel far from their home location. Figure~\ref{fig:map360916} shows agent 360916, who visits a large number of workplaces, restaurants, and recreational sites around their home location, but does not visit many of them regularly.

\begin{figure}[b]
\vspace*{-1ex}
    \centering
    \includegraphics[width=0.9\linewidth]{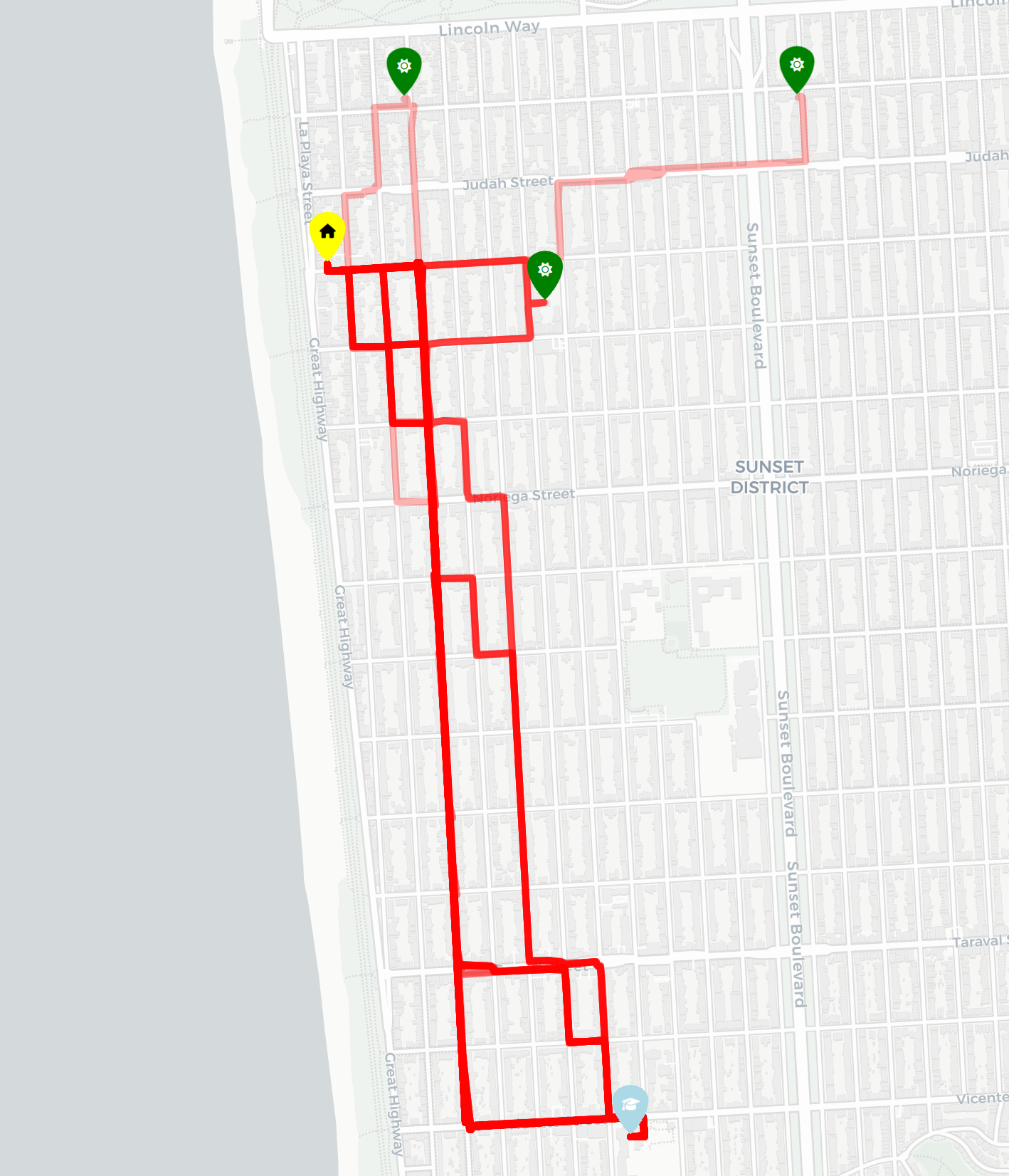}
    \caption{Agent 49270 (student). Staypoint colors indicate location type and follow calendar plot key.}
    \label{fig:map49270}
\end{figure}

\begin{figure}[b]
    \centering
    \includegraphics[width=0.9\linewidth]{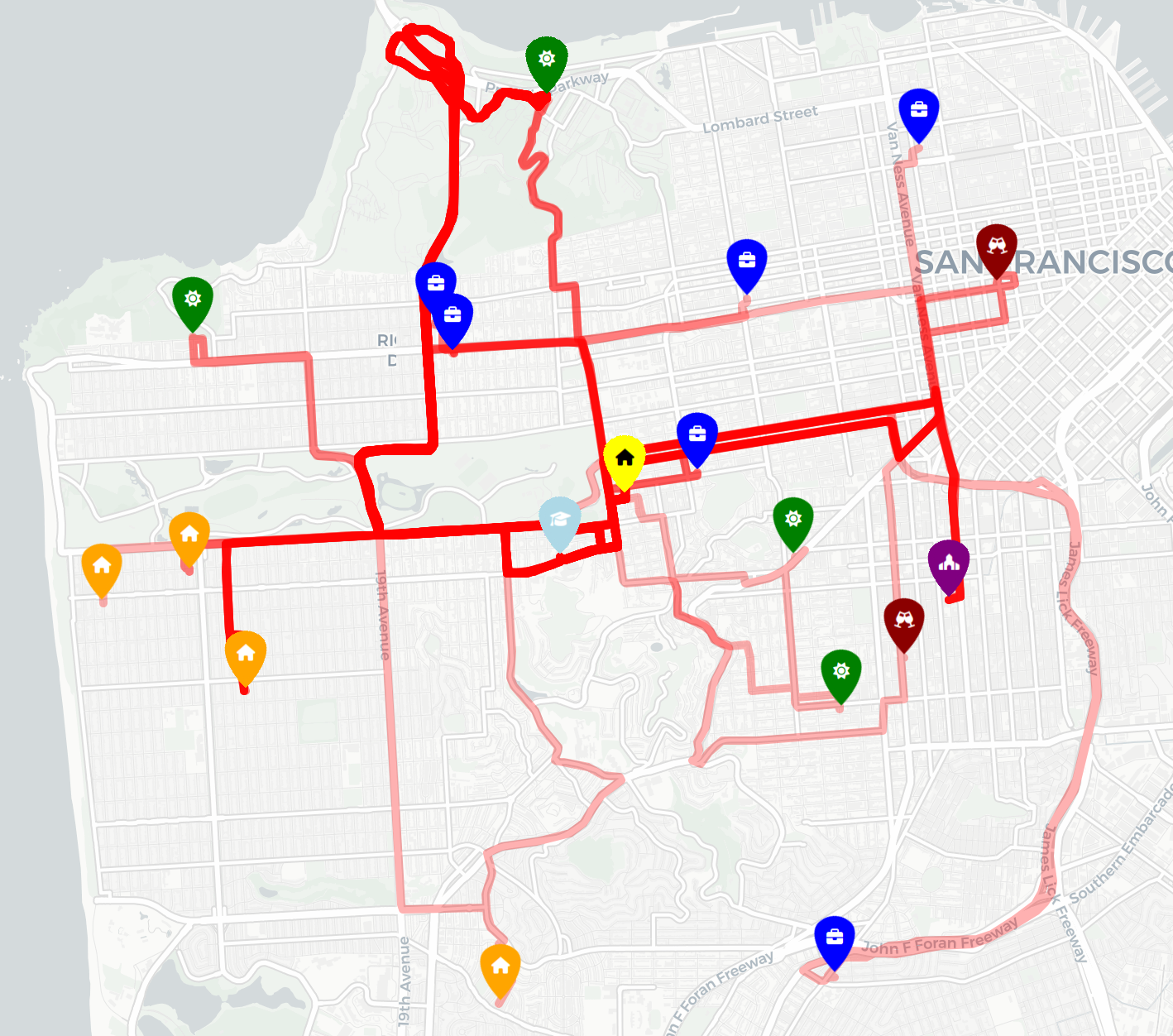}
    \caption{Agent 66982 (worker). Staypoint colors indicate location type and follow calendar plot key.}
    \label{fig:map66982}
\end{figure}

Figure~\ref{fig:map49270} shows agent 49270, who displays very simple patterns of life: they regularly travel between their home and a school location, occasionally (but rarely) visiting recreational locations as well. Figure~\ref{fig:map66982} shows agent 66982, who visits several different location types, all centered around their home location, including workplace, recreational, and a religious location. We see they make stops at a nearby school, which based on their calendar plot in Figure~\ref{fig:cal66982}, 
appear to be dropoff and pickups for a child.

\begin{figure}[b]
    \centering
    \includegraphics[width=0.9\linewidth]{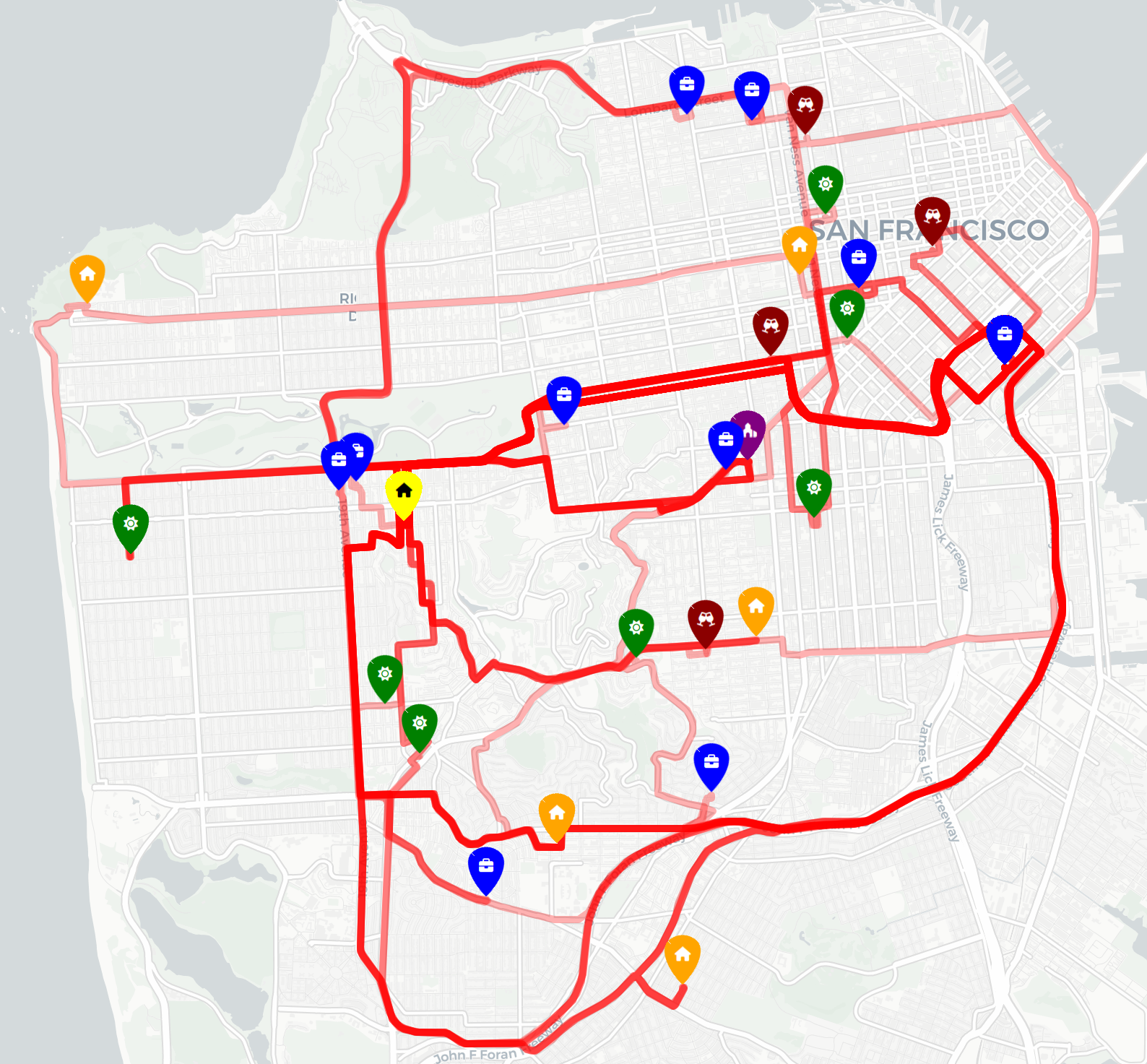}
    \caption{Agent 439557 (worker). Staypoint colors indicate location type and follow calendar plot key.}
    \label{fig:map439557}
\end{figure}

\begin{figure}[b]
    \centering
    \includegraphics[width=0.9\linewidth]{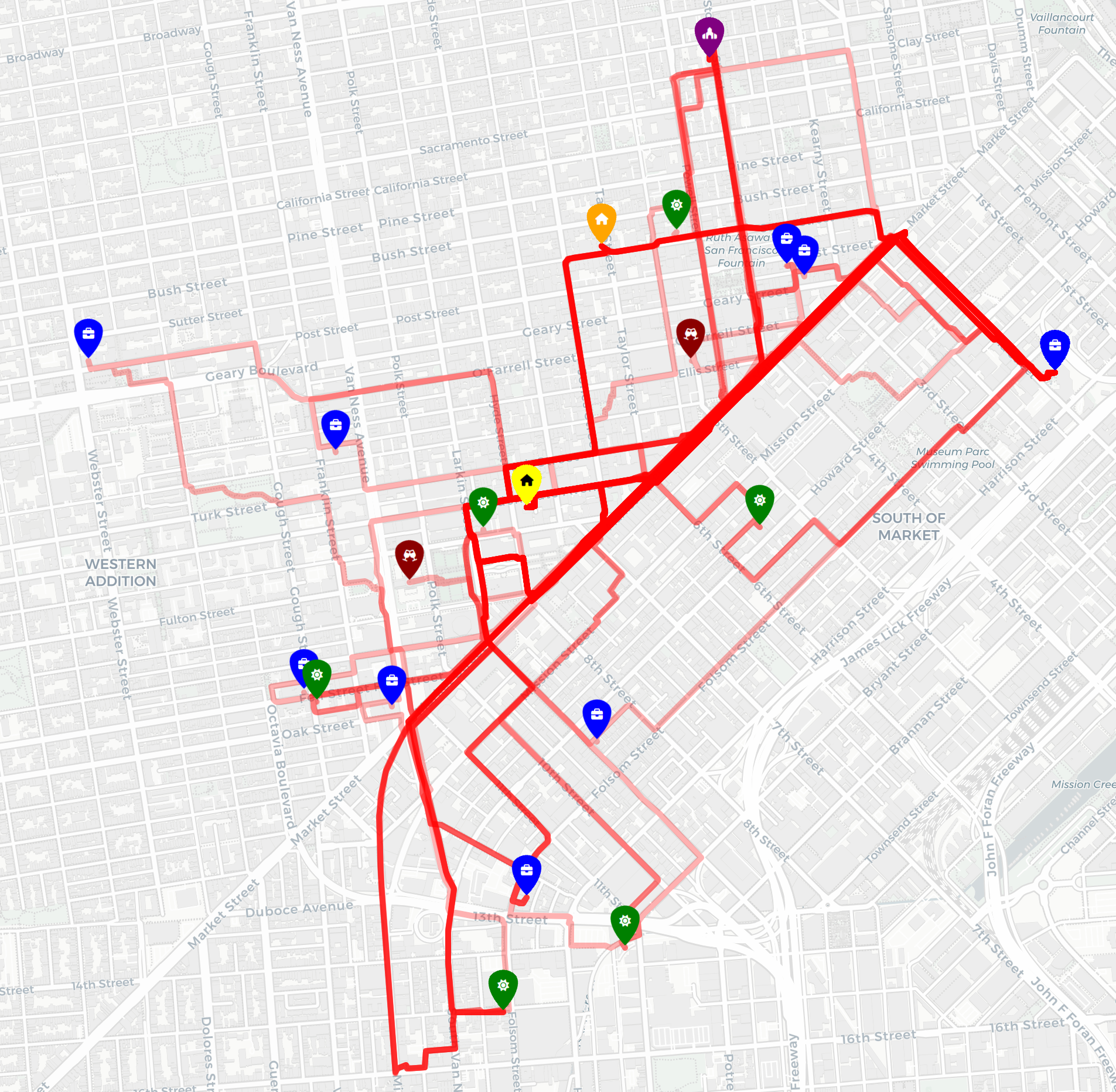}
    \caption{Agent 479018 (worker). Staypoint colors indicate location type and follow calendar plot key.}
    \label{fig:map479018}
\end{figure}

Figure~\ref{fig:map439557} shows agent 439557, who visits many location types with some far from home, suggesting greater activity than most agents shown here. Figure~\ref{fig:map479018} shows agent 479018, who visits fewer unique locations and tends to stay closer to home. Both agents regularly attend a religious location as well.

\begin{figure}[htbp]
    \centering
    \includegraphics[width=\linewidth]{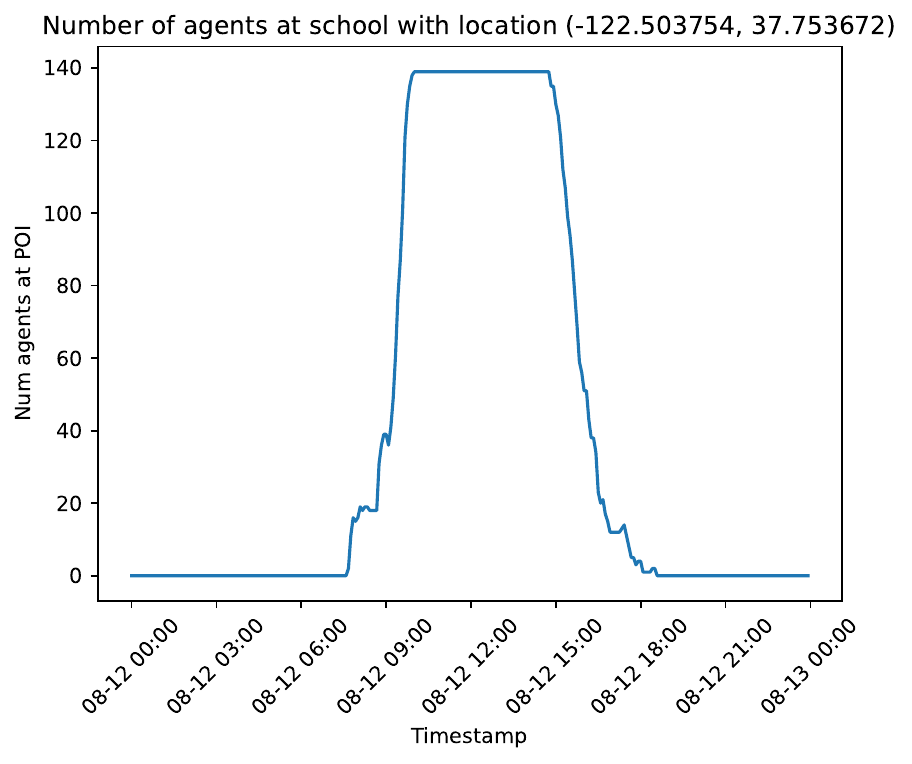}
    \caption{Number of agents present at a given school during one day of simulation.}
    \label{fig:one_day}
\end{figure}

\begin{figure}[htbp]
    \centering
    \includegraphics[width=\linewidth]{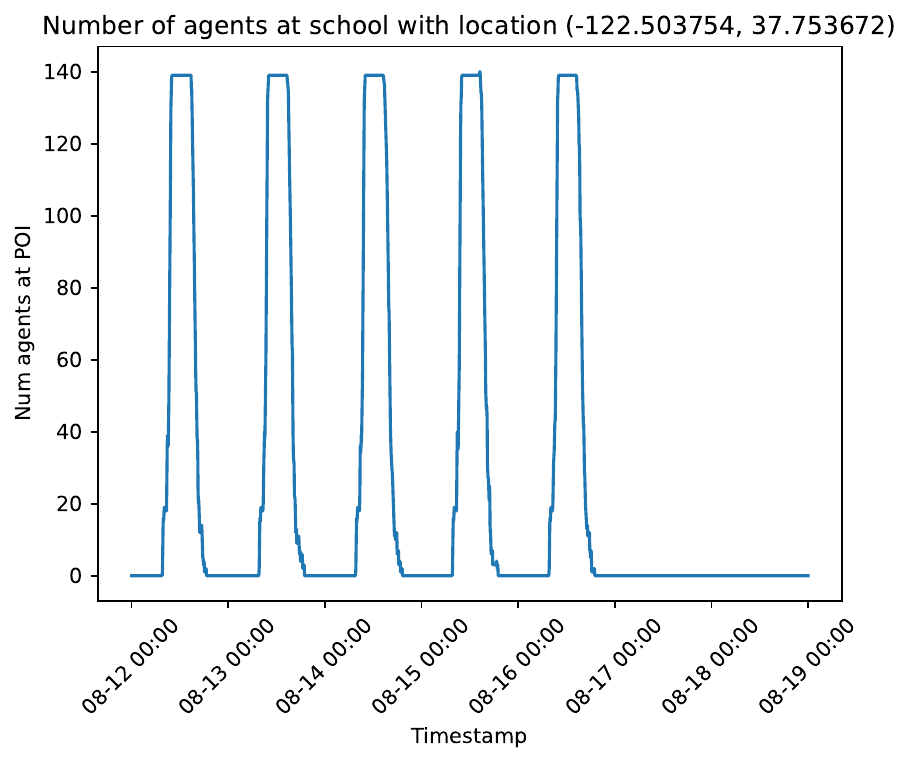}
    \caption{Number of agents present at a given school during one week of simulation.}
    \label{fig:one_week}
\end{figure}

On the other hand, Figures~\ref{fig:one_day} and~\ref{fig:one_week} show the number of agents at a particular school location throughout time in the simulation period. Figure~\ref{fig:one_day} shows a single day, and we observe that agents tend to arrive around 9 AM and depart around 5 PM, with some variance. We also see small peaks and troughs during the arrival departure periods as some parents have brief stays for pickup and dropoff. In Figure~\ref{fig:one_week}, we see the same school location for a week, and see similar patterns on weekdays, but no attendance on weekends, when school is not in session.

\section{Conclusion}

SF-LIFE represents a significant contribution to the spatial computing and transportation research communities, providing a massive-scale, high-frequency, noise-free, and accessible movement dataset for the San Francisco Bay Area. The dataset's combination of (1) realistic simulation of needs-based human behavior, (2) kinematic simulation of 1Hz frequency mobility, (3) labeled agent activity agendas, (4) synthetic population  demographic data, and (5) OSM environment data, make this dataset an ideal resource for transportation analytics, machine learning research, and urban computing applications, especially in cases where research would like to scale their methods to datasets much larger than publicly available dataset. Future work will focus on expanding the dataset to include additional time periods, geographic regions, and transportation modes. We also plan to develop companion tools and benchmarks to facilitate research using the dataset.

\begin{acks}
Supported by the Intelligence Advanced Research Projects Activity (IARPA) via Department of Interior/ Interior Business Center (DOI/IBC) contract number 140D0423C0025. The U.S. Government is authorized to reproduce and distribute reprints for governmental purposes notwithstanding any copyright annotation thereon. Disclaimer: The views and conclusions contained herein are those of the authors and should not be interpreted as necessarily representing the official policies or endorsements, either expressed or implied, of IARPA, DOI/IBC, or the U.S. Government.
\end{acks}

\bibliographystyle{ACM-Reference-Format}
\bibliography{references}


\end{document}